\documentclass[sigconf, nonacm]{acmart}

\AtBeginDocument{  \providecommand\BibTeX{{    \normalfont B\kern-0.5em{\scshape i\kern-0.25em b}\kern-0.8em\TeX}}}

\usepackage{makecell}
\usepackage{weiwAlgorithm}

\usepackage{graphicx}
\usepackage{balance}  
\usepackage{lipsum}
\usepackage{amsmath}
\usepackage{physics}
\usepackage{subfigure}
\usepackage{tabularx}
\usepackage{color}
\usepackage{colortbl}
\usepackage{float}
\usepackage{listings}
\usepackage{multirow}
\usepackage[normalem]{ulem}
 \usepackage{longtable}
\usepackage{enumitem}
\usepackage{multirow}
\usepackage{booktabs}
\usepackage{capt-of}
\usepackage{pifont}
\usepackage{ulem}
\usepackage{soul}
\usepackage{cancel}
\usepackage{bm}
\usepackage{url}
\usepackage{caption}

\def\BibTeX{{\rm B\kern-.05em{\sc i\kern-.025em b}\kern-.08em
    T\kern-.1667em\lower.7ex\hbox{E}\kern-.125emX}}

 \graphicspath{{./Graph/}, {./Fig/}, {./Legend/}}

\long\def\comment#1{}

\newcounter{example}[section]
\renewcommand{\theexample}{\nthesection.\arabic{example}}
\newenvironment{example}{
     \refstepcounter{example}
     {\vspace{1ex} \noindent\bf  Example  \theexample:}}{
     \vspace{1ex}} 

\newcounter{definition}[section]
\renewcommand{\thedefinition}{\nthesection.\arabic{definition}}

\newcounter{theorem}[section]
\renewcommand{\thetheorem}{\nthesection.\arabic{theorem}}
\newenvironment{theorem}{\begin{em}
        \refstepcounter{theorem}
        {\vspace{1ex} \noindent\bf  Theorem  \thetheorem:}}{
        \end{em}\vspace{1ex}} 

\newcounter{lemma}[section]
\renewcommand{\thelemma}{\nthesection.\arabic{lemma}}

\newcounter{corollary}[section]
\renewcommand{\thecorollary}{\nthesection.\arabic{corollary}}

\newcounter{proposition}[section]
\renewcommand{\theproposition}{\nthesection.\arabic{proposition}}

\newcounter{remark}[section]
\renewcommand{\theremark}{\nthesection.\arabic{remark}}

\newcommand{\proofsketch}{\noindent{\bf Proof Sketch: }}

\newcommand{\nthesection}{\arabic{section}}

\newcommand{\eop}{\hspace*{\fill}\mbox{$\Box$}\vspace*{1ex}}

\newcommand{\stitle}[1]{\vspace{1mm} \noindent{\bf #1}}

\newcommand{\green}[1]{\textcolor{green}{}}

\newcommand{\kw}[1]{{\ensuremath {\mathsf{#1}}}\xspace}
\newcommand{\kwnospace}[1]{{\ensuremath {\mathsf{#1}}}}

\newcommand{\dist}{\delta} 

\newcommand\bigutimes{\mathop{\ooalign{$\bigcup$\cr%
   \hfil\raise0.36ex\hbox{$\scriptscriptstyle\boldsymbol{\times}$}\hfil\cr}}}
\newcommand\bigumius{\mathop{\ooalign{$\bigcup$\cr%
   \hfil\raise0.36ex\hbox{$\scriptscriptstyle\boldsymbol{-}$}\hfil\cr}}}

\newcommand{\R}{\mathbb{R}\xspace} 

\newcommand{\SIFT}{\kw{SIFT}}
\newcommand{\GIST}{\kw{GIST}}

\newcommand{\yaleb}{\kw{YaleB}}
\newcommand{\ffhq}{\kw{FFHQ}}
\newcommand{\celebA}{\kw{CelebA}}
\newcommand{\multipie}{\kwnospace{Multi}-\kw{PIE}}

\newcommand{\deepcolor}{\cellcolor[rgb]{0.8745, 0.8235, 0.7137}}
\newcommand{\lightcolor}{ \cellcolor[rgb]{0.9882,0.9373,0.8196}}

\newcommand{\Pset}{\mathcal{P}}

\newcommand{\Theory}{\kw{TheoryLP}}
\newcommand{\Lift}{\kw{Lift}}
\newcommand{\DirSearch}{\kw{DirectSearch}}
\newcommand{\Brute}{\kw{BruteScan}}
\newcommand{\ours}{\kw{LinePG}}
\newcommand{\Epair}{E_{\mathrm{pair}}}
\newcommand{\Ereg}{E_{\mathrm{reg}}}

\newcommand{\Mpair}{M_{\mathrm{pair}}}
\DeclareMathOperator{\sveczero}{svec_0}

\begin{document}

\title{Fast Approximate Line Nearest Neighbor Search}

\author{Jiadong Xie}
\orcid{0000-0003-4535-8359}
\affiliation{%
  \institution{The Chinese University of Hong Kong}
  \country{}
}
\email{jdxie@se.cuhk.edu.hk}

\author{Jeffrey Xu Yu}
\orcid{0000-0002-9738-827X}
\affiliation{%
  \institution{The Hong Kong University of Science and Technology (Guangzhou)}
  \country{}
}
\email{jeffreyxuyu@hkust-gz.edu.cn}

\author{Yingfan Liu}
\orcid{0000-0002-3743-5249}
\affiliation{%
  \institution{Xidian University}
  \country{}
}
\email{liuyingfan@xidian.edu.cn}

\begin{abstract}
Approximate nearest neighbor (ANN) search, which retrieves vectors close to a given query vector, is a fundamental operation in vector databases. However, emerging applications, such as selecting source images for photo-based animation and video synthesis, may require matching content while allowing a query-specified aspect contrast, such as an illumination change, to vary; a single query point does not directly express this requirement. In this paper, we study approximate line nearest neighbor (LANN) search, where an infinite query line removes the distance contribution along a specified direction while database records remain points. Prior approaches focus on top-$1$ search only for LANN and can incur high construction costs. Converting LANN into ordinary point ANN search through dimension lifting substantially increases index construction and query costs, while direct search on a proximity graph built for ordinary point ANN search can miss distinct regions close to a line. To address these issues, we propose \ours, which enhances an existing proximity graph with pair edges to facilitate navigation between separated near-line regions. To find these pair edges efficiently, we first generate region pairs with spatial or graph separation, and near a line. We then score edges between the selected region pairs to guide the point endpoint selections. We conduct extensive performance studies and confirm the accuracy and efficiency of \ours in comparison with the baseline approaches.
\end{abstract}


\maketitle

\section{Introduction}

Approximate nearest neighbor (ANN) search is a fundamental operation in vector databases~\cite{vector2026tutorial,WangXY021,WangYGJXLWGLXYY21,ZhangXCSXCCH00Y23,YinGBCL25}. Given a query vector, it retrieves data vectors that are close under a specified distance. Many approaches have been proposed for this ANN query, including proximity graph methods that navigate between nearby data points~\cite{taumg,LiYL26,dpg,NEURIPS2019_09853c7f,WangWCWPW24,LuT25,AziziEP25,ManoharSBD0S024,ZhaoTHTXZZ26}, which are among the state-of-the-art approaches for ANN search. 
However, emerging applications, such as visual media creation, introduce additional requirements for query processing.
Google's Picasa included a Face Movies feature based on Exploring Photobios, which selects and orders photographs of the same person to create face animations~\cite{Kemelmacher-ShlizermanSGS11}.
Both Runway's Gen-4 References~\cite{runway2025gen4references} and photo-based time-lapse mining~\cite{Seitz15} need to maintain subject or scene consistency while handling illumination differences in video-creation workflows.
These workflows motivate retrieving content-consistent objects despite query-specified aspect changes.
For example, when assembling a same-subject, same-pose face sequence across illumination changes for video generation, a user needs images that match the subject and pose while allowing lighting conditions to change.
An ordinary ANN uses a query point for search, it does not explicitly specify which aspect variation should be disregarded, and can thus favor similar lighting over the characteristics the user wishes to preserve.

\begin{figure}
  \includegraphics[width=\columnwidth]{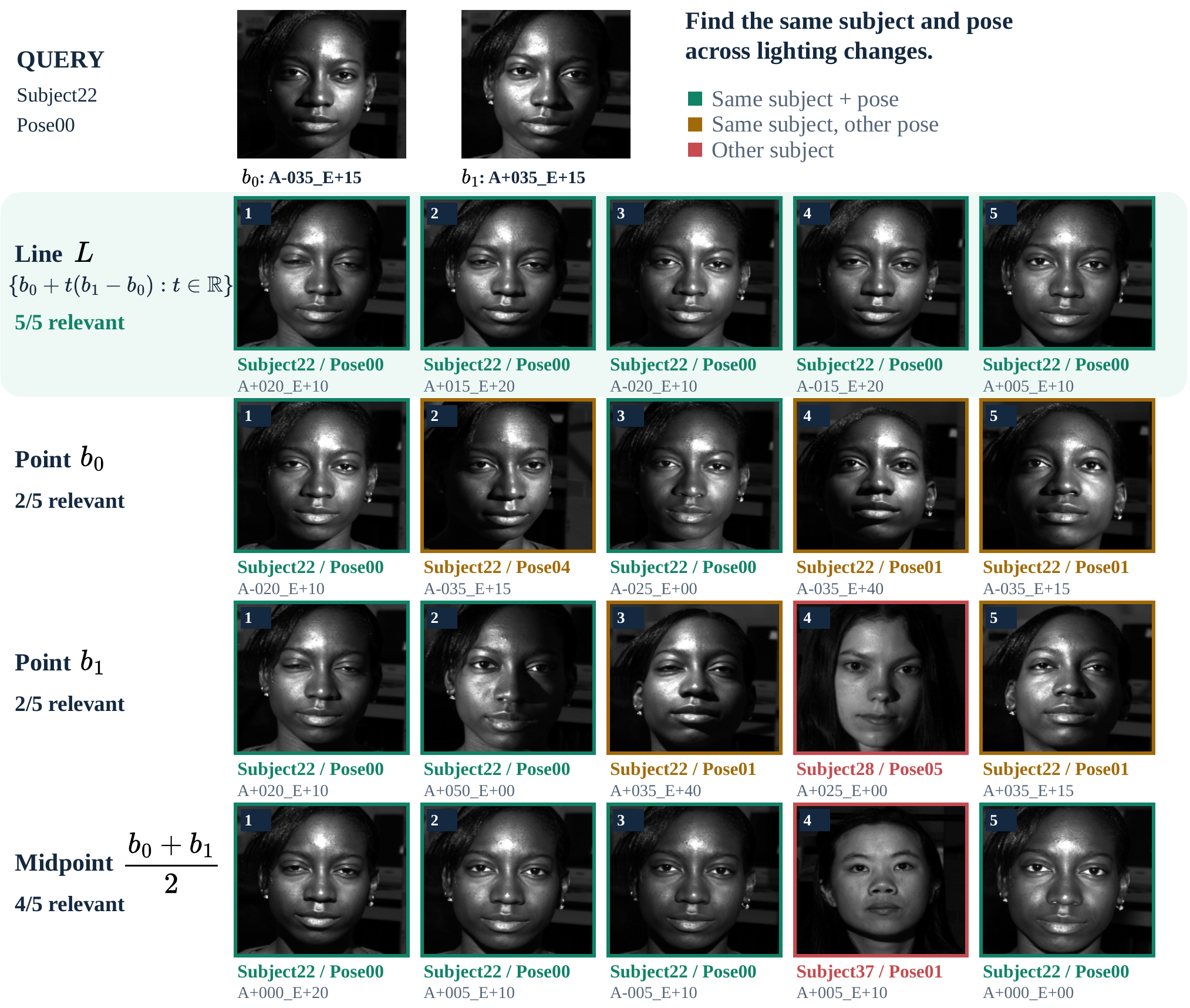}
  \vspace*{-5mm}
  \caption{A Case Study on \yaleb Dataset}
  \label{fig:case-study}
  \vspace*{-4mm}
\end{figure}

A directional model of the allowed aspect change provides a way to express this retrieval requirement.
Image inversion maps real images to latent representations suitable for editing~\cite{TovANPC21}, while latent-direction studies identify
approximately linear directions for selected facial-attribute
changes~\cite{ShenGTZ20}.
A similar geometric formulation arises in query-side tangent-distance methods~\cite{SimardCD92,KeysersDTN00}, where a one-parameter image transformation is locally approximated by an affine line, and similarity to a fixed database vector is measured by its minimum distance to this line.
These findings motivate representing a specified aspect change by an affine line $z+tu$, where $z$ is an image code, $u$ specifies the direction, and $t$ controls the amount of change.
For retrieval, the objective is to find existing images compatible with the specified variation; we therefore consider an infinite line query $z+\mathbb R u$, which ranks database vectors by their minimum distance to the line without fixing the amount of change $t$.
In the illumination example, two observations of the same subject and pose under different lighting conditions provide a query-specific secant direction through the difference between their embeddings.
Fig.~\ref{fig:case-study} illustrates this retrieval requirement on \yaleb~\cite{GeorghiadesBK01,LeeHK05}. The two query images show subject22 in pose00, with illumination azimuths of $-35^\circ$ and $+35^\circ$, both at an elevation of $15^\circ$.
Using their 128-dimensional embeddings as the line endpoints, we compare exact top-5 retrieval over the same 16{,}126 candidate images, excluding both query images.
Line retrieval returns five images matching both the query subject and pose, compared with two for either endpoint query and four for their midpoint.
The line results span different lighting conditions, whereas the point-based alternatives also return images of other subjects or poses.
In the figure, A/E denote illumination azimuth/elevation in degrees.

Motivated by these query requirements, in this paper, we study {line nearest neighbor search}. Given a dataset $\Pset\subseteq\R^d$, a query is an infinite affine line $L=\{b_0+t(b_1-b_0):t\in\R\}$ through distinct endpoints $b_0,b_1$. The search retrieves $k$ nearest data points of the query line $L$, i.e., under the distance $\dist(p,L)=\min_{t\in\R}\norm{p-b_0-t(b_1-b_0)}.$
The database continues to store one vector per record. The line specifies which component of the difference is not penalized. 
Due to the curse of dimensionality~\cite{IndykM98}, the exact line nearest neighbor search is computationally intensive. We focus on the approximate line nearest neighbor search (LANN) in this paper, which has not been well studied.
If all queries shared one direction, a fixed projection followed by ordinary point ANN would suffice. We consider varying directions and offsets, as in the endpoint-defined workloads, so one fixed projection cannot serve all queries.

In the literature, existing approaches for LANN is theoretical studies for top-$1$ search~\cite{AndoniIKN09,MulzerNSS15}. They decompose vector data into dense near-line groups and a sparse residual set, but literal group construction is expensive, higher than quadratic in dataset size. Alternatively, existing point-to-hyperplane nearest neighbor search~\cite{HuangLT21} provides an insight to solve LANN problem by lifting the vector dimension, and then converts the query into a point query in a higher-dimensional Euclidean space. However, we prove that the lowest dimension we need for LANN problem to solve in such a way is $\Theta(d^2)$, which significantly increases the index construction and search cost due to the much larger dimension distance computation that is needed. Another idea is to reuse a proximity graph built for point ANN search and replace its point-to-point comparisons with point-to-line distances.
The difficulty is that points close to one line need not be close to each other. They can occupy several regions separated along the line, while paths between those regions pass through points farther from the line. Hence, a search may find enough candidates in one region and stop before entering another region with better results.

In this paper, we propose \ours, which constructs a line-enhanced proximity graph from an existing proximity graph built for point ANN search. 
The main contributions of this work are summarized below. 
\ding{202} We analyze why existing point-search structures can miss multiple near-line regions, motivating pair edges that connect separated regions close to a common line. These edges are used to enhance the original proximity graph  to facilitate navigation between separated near-line regions.
\ding{203} To find these pair edges efficiently and avoid examining all point pairs, we propose first finding promising pairs of regions.
We introduce a balanced hierarchy that considers both vector locations and proximity graph connections to find spatially separated candidate regions, together with a quotient graph that identifies region pairs without short connecting paths.
Next, a geometric score combines separation with support near a representative line to rank candidate region pairs.
\ding{204} We propose an edge-scoring mechanism to select pair edges, turning these region pairs into useful edges in the graph for navigation without increasing the total degree limit.
It selects endpoints whose connecting direction is supported by multiple points in the target region and discourages directions redundant with existing or previously selected edges.
The selected pair edges enhance the proximity graph, replacing retained original edges only when necessary to satisfy the total degree limit.
\ding{205} We further design a search algorithm for the line-enhanced proximity graph that uses selected pair edges as additional entrances into promising regions while avoiding redundant entrances into regions already being explored.
\ding{206} Experiments on six real-world datasets show that \ours achieves $1.44$--$4.23$x QPS of direct search on the original proximity graph when both methods reach at least $99\%$ recall. On other datasets, \ours can reach higher recall than $99\%$, while direct search can not. Compared with dimension lifting and existing theoretical methods, \ours achieves QPS speedups of up to $976.17$x and $1{,}665.63$x, respectively, with much smaller index cost, at $100\%$ recall on the $N=512$ subsets.

The paper is organized as follows. Sections~\ref{sec:pre} and~\ref{sec:limit} define the problem and discuss existing approaches. Section~\ref{sec:overview} presents our design. Sections~\ref{sec:cluster} and~\ref{sec:edges} give the region construction and candidate region pair selection algorithms. Section~\ref{sec:search} describes search, followed by experiments in Section~\ref{sec:exp}. Section~\ref{sec:con} concludes this work.

\section{Preliminaries and Problem Statement}
\label{sec:pre}

Given a set of points $\Pset=\{p_1,\ldots,p_N\}$, in Euclidean $d$-dimensional space, denoted as $\R^d$, for $d\ge 2$, we represent a nondegenerate \textbf{line} query by $ L(a,u)=\{a+tu:t\in\R\},$ where $\norm{u}=1$ representing direction, and $a^\top u=0$.
The orthogonality condition makes $a$ the point on $L$ closest to the origin.
Two supplied endpoints $b_0,b_1$, with $b_0\neq b_1$, yield $u=(b_1-b_0)/\norm{b_1-b_0}$ and $a=b_0-(b_0^\top u)u$. Reversing $u$ does not change $L$.
{Fig.~\ref{fig:notations} summarizes the main geometric notation used throughout the paper.}
Let $H_u=I-u u^\top$ be the orthogonal projector onto the complement of $u$. The distance between a point $p\in \R^d$ and a line $L$ is $\dist(p,L)=\norm{H_u(p-a)}$.
We use $\dist^2(p,L)$ for ranking: $\dist^2(p,L)=\norm{p}^2-2p^\top a+\norm{a}^2-(p^\top u)^2$, where $\norm{p}^2$ can be precomputed for every point in $\Pset$ and $\norm{a}^2$ can be computed once before searching.
For a line $L$ and $r>0$, the set $S_r(L)=\{p\in\Pset:\dist(p,L)\le r\}$ is the dataset portion of an infinite cylinder, as illustrated in Fig.~\ref{fig:motivation-linePG}, $S_r(L)=\{p_4,p_5,p_6,p_7,p_8,p_9\}$.

\stitle{Line ANN Search:}
Given a query line $L$ and a positive integer $k\le N$, line nearest neighbor search aims to find the $k$ points in $\Pset$ closest to $L$. We denote the set of these points as the line $k$-nearest neighbors, $\mathrm{NN}_k(L)=\{p_1^*,\ldots,p_k^*\}$, and their distances as $r_1^*(L)\le r_2^*(L)\le\ldots\le r_k^*(L)$. Since exact search is expensive, we study \textbf{approximate line nearest neighbor (LANN) search} in this paper. We denote the returned set by $\widehat{\mathrm{NN}}_k(L)=\{\hat p_1,\ldots,\hat p_k\}$, and its distances by $\widehat r_1\le \widehat r_2\le\ldots\le \widehat r_k$.
Following existing point ANN studies~\cite{taumg,nsg,hnsw,abs-2410-01231}, we use recall to measure the retrieval quality of approximate LANN search: $ \mathrm{Recall@}k={|\widehat{\mathrm{NN}}_k(L)\cap\mathrm{NN}_k(L)|}/{k}.$

\stitle{Point ANN Search and Proximity Graphs:}
For a query point $q\in\R^d$, conventional point ANN search retrieves $k$ approximate nearest neighbors under $\dist(p,q)=\norm{p-q}$. Graph-based methods~\cite{nsg,hnsw,taumg,ALMG,LiYL26,AziziEP25}, which are among the state-of-the-art approaches for point ANN search, support this operation using a proximity graph $G=(V,E)$. Each node in $V$ represents one database point, and an edge $(p_i,p_j)\in E$ connects close points selected using point-to-point distances. We use the same symbol for a point and its graph node, and write $N_G(p_i)=\{p_j:(p_i,p_j)\in E\}$ for its out-neighbor list. A degree parameter $M$ controls the list size. The selected edges need not be exactly the $M$ nearest-neighbor edges: pruning and additional navigation edges also affect the topology~\cite{nsg,hnsw}.

Point ANN search performs beam search on the proximity graph; it starts from one or more entry nodes and maintains a candidate pool containing at most $w\ge k$ nodes. It repeatedly expands the closest unexpanded candidate, evaluates its previously unseen out-neighbors, and retains the $w$ closest candidates encountered. Search stops when no retained candidate remains to be expanded and returns the $k$ closest retained points.

\section{Limitations of Existing Approaches}
\label{sec:limit}

\begin{figure}[t]
  \centering
  \begin{minipage}{0.48\linewidth}
  \includegraphics[width=0.94\linewidth]{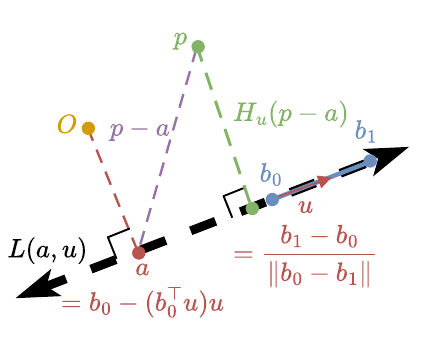}
  \caption{An illustration of main geometric notations}
  \label{fig:notations}
  \end{minipage}
  \hspace{1mm}
  \begin{minipage}{0.48\linewidth}
  \includegraphics[width=\linewidth]{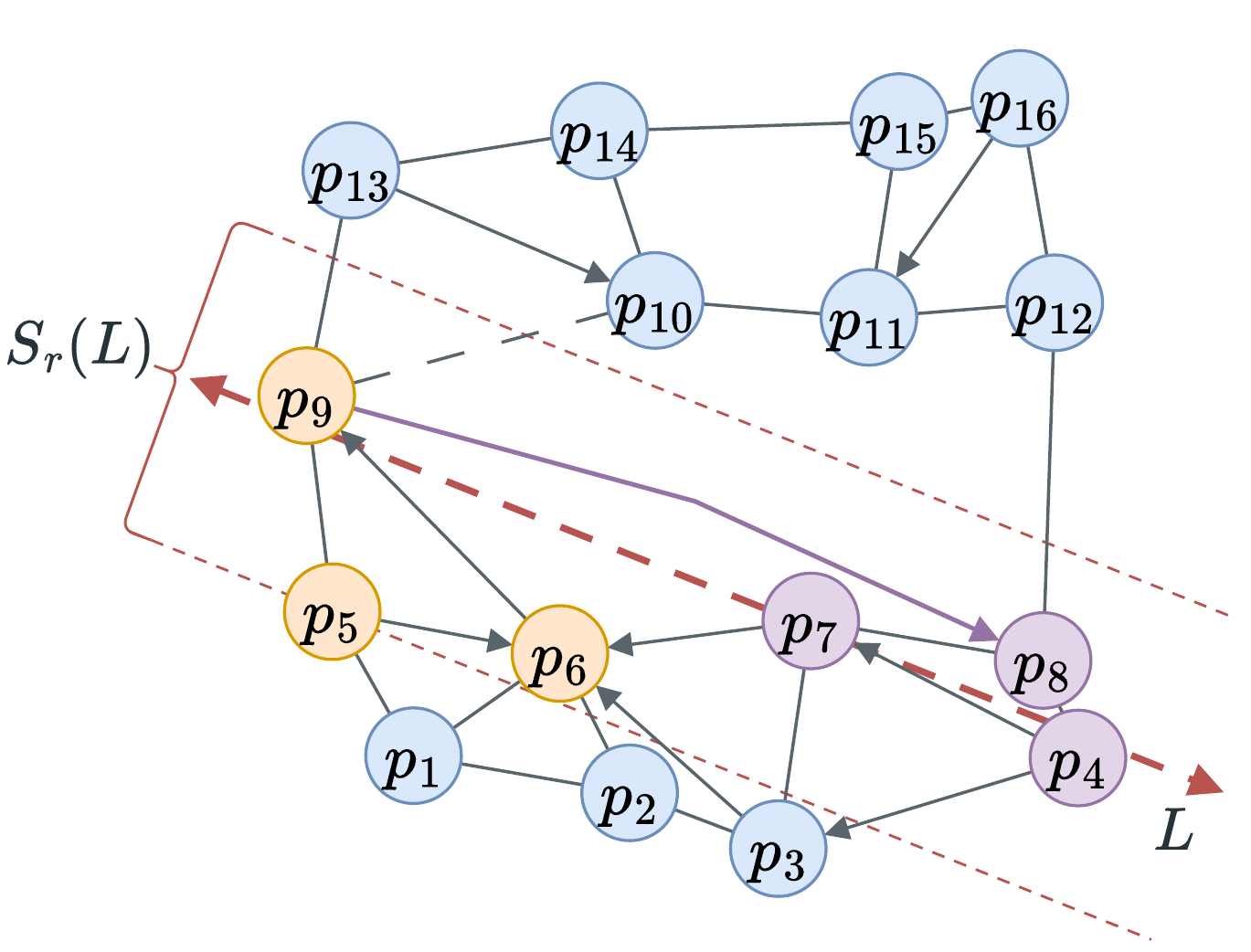}
  \caption{An example of a proximity graph}
  \label{fig:motivation-linePG}
  \end{minipage}
  \vspace{-4mm}
\end{figure}

In this section, we discuss the limitations of existing approaches to support LANN search.

\stitle{Provable Line ANN Search:}
Andoni et al.~\cite{AndoniIKN09} study LANN search with $k=1$ from a theoretical perspective; we denote their approach as \Theory.
Its objective is to return, with constant success probability, a point whose distance to the query line is within a factor $(1+\epsilon)$ of the optimum. Later work by Mulzer et al.~\cite{MulzerNSS15} supports fixed-dimensional affine-flat queries, including lines, with improved space bounds. Both guarantees concern one returned neighbor.

At a high level, \Theory decomposes the dataset into \emph{dense near-line groups} and a \emph{sparse residual set}.
Dense near-line groups are selected iteratively. In each round, \Theory examines axes through pairs of remaining data points, and chooses the axis with the smallest $m$-th point-to-axis distance, where $m$ is of order $\sqrt N$.
The points within this radius form a dense group, for which a {Cylinder-DS} is constructed to reduce line ANN search inside a dense cylinder to a small number of point ANN queries.
For the points not handled by sufficiently thin cylinders, i.e., points in the sparse residual set, \Theory uses a {Projection-DS}, which randomly projects the data into a constant-dimensional space and reduces the resulting low-dimensional line nearest-neighbor problem to a geometric ray-shooting problem following Magen~\cite{Magen07}.

At query time, \Theory first uses the Projection-DS over the full dataset to obtain a coarse neighbor $\tilde p$ and estimate the relevant distance scale.
It then queries the Cylinder-DS structures whose radii are sufficiently small relative to this estimate, while the remaining points are handled by a Projection-DS.

\noindent\underline{\bf Issues:}
First, \Theory's native result and approximation guarantee are designed for the $k=1$ case, rather than the $k\geq 1$ objective considered here. Second, the literal cylinder selection is expensive in \Theory. If round $i$ contains $n_i$ active points, it evaluates $\binom{n_i}{2}$ axes against those points. With linear-time selection of each $m$-th distance, the resulting work is $O\!\left(d\sum_i n_i^3\right)=O(dN^{7/2})$ for $m=\Theta(\sqrt N)$, for this selection procedure alone. This results in prohibitively high construction overhead.
In our implementation, we extend \Theory to support $k\geq 1$ by maintaining the $k$ nearest candidates among all data points evaluated during the search. Nevertheless, {as shown in Exps.~1 and 2, its search efficiency remains substantially worse than that of the other approaches, even if its construction cost is much higher than others.}

\stitle{Direct Search on a Proximity Graph:}
A straightforward approach reuses a proximity graph built for point ANN search, as described in Section~\ref{sec:pre}, but replaces the distance between two points with $\dist(p,L)$, the distance between a line and a point.

\noindent\underline{\bf Issues:}
Closeness to the same line does not imply closeness to each other.
For a line $L(a,u)$, let $\tau_L(p)=p^\top u$ be a point's coordinate along $L$, and let $e_L(p)=p-a-\tau_L(p)u$ be its orthogonal residual. For any two points, $\norm{p-q}^2 =\bigl(\tau_L(p)-\tau_L(q)\bigr)^2+\norm{e_L(p)-e_L(q)}^2.$
Points in $S_r(L)$ can thus be arbitrarily far apart along $L$, even though $\|e_L(p)-e_L(q)\|\le 2r$.
However, in a proximity graph, edges connect nearby points~\cite{WangXY021,vector2026tutorial}. Hence, a proximity graph may connect such regions only through points farther from $L$.

\begin{example}
\label{ex:direct}
Fig.~\ref{fig:motivation-linePG} shows a proximity graph of $\Pset=\{p_1,\ldots,p_{16}\}$ with out-degree limit 3 (including edge $(p_9,p_{10})$ without $p_9\to p_8$), and unarrowed edges denote both directions. Consider the line $L$ and the set $S_r(L)$, with $p_9$ as the entry point. The points in $S_r(L)$ form two regions in the induced graph: $R_1=\{p_5,p_6,p_9\}$ and $R_2=\{p_4,p_7,p_8\}$. All directed paths from $R_1$ to $R_2$ pass through a point outside $S_r(L)$, since there is no edge from $p_6$ to $p_7$. The beam search with $w=3$ terminates with $\{p_9,p_6,p_5\}$, since expanding $p_6$ does not provide a better continuation, although $3$-NN of $L$ are $p_7,p_4,p_9$.
\end{example}

Hence, beam search on a proximity graph may terminate when finding enough candidates in one region, leaving better points in another region undiscovered.

\stitle{Reduction via Quadratic Lifting:}
Inspired by the asymmetric quadratic transformation for point-to-hyperplane search~\cite{HuangLT21}, we consider converting LANN search into point ANN search via dimension lifting.
Specifically, we first apply the same translation and uniform scaling to the dataset and query line so that every data point has norm at most one.
This preserves point-to-line distance rankings.
For simplicity, we reuse $p,a,u$ for normalized coordinates, with the transformed line query expressed in canonical form.

Let $S(p)=pp^\top-\norm{p}^2 I/d$, $T(u)=uu^\top-I/d$, and $\gamma=d/(d-1)$. Write $\sveczero$ for Frobenius-orthonormal coordinates of traceless symmetric matrices; this block has $d(d+1)/2-1$ entries.
We map each data point $p=(p_1,\ldots,p_d)$ and query line $L(a,u)$ to $ X(p)=\left(\sveczero(S(p)),p,   \sqrt{1-\tfrac{d-1}{d}\norm{p}^4}\right),$ and $ Y(L)=\left(\tfrac\gamma2\sveczero(T(u)),\gamma a,0\right).$
Both vectors have $D_\star=d(d+1)/2+d$ dimensions.
Since $\norm{X(p)}^2=1+\norm{p}^2$ and $\langle S(p),T(u)\rangle_F=(p^\top u)^2-\norm{p}^2/d$, their squared Euclidean distance satisfies $\norm{X(p)-Y(L)}^2=1+\norm{Y(L)}^2+\gamma\bigl(\dist^2(p,L)-\norm{a}^2\bigr).$
For a fixed query, $\gamma>0$ preserves the exact nearest-neighbor ordering. We can therefore build a point ANN index over $X(p)$ and query it with $Y(L)$. The following theorem shows this construction uses the minimum dimension in the stated exact-reduction model.

\begin{theorem}
\label{thm:lift-optimal}
Let $d\ge2$ and let $\Omega\subseteq\R^d$ have nonempty interior. Any fixed maps $\Phi:\Omega\to\R^D$ and $\Psi(L)\in\R^D$ satisfying $\norm{\Phi(p)-\Psi(L)}^2=c(L)\dist^2(p,L)+\eta(L), c(L)>0,$ for all $p\in\Omega$ and all affine lines $L$ require $D\ge D_\star=d(d+1)/2+d$. The construction attains this bound on the unit ball.
\end{theorem}

\proofsketch
\sloppy
Modulo constants, the squared line-distance functions span the $D_\star$-dimensional space $\mathcal V$ of quadratic and linear polynomials: for $d\ge2$, the matrices $I-uu^\top$ span all symmetric matrices (their sum over coordinate directions is $(d-1)I$), and varying $a\perp u$ supplies all linear terms. Expanding $\norm{\Phi(p)-\Psi(L)}^2=c(L)\dist^2(p,L)+\eta(L)$, with $c(L)\ne0$, gives $\mathcal V\subseteq\mathcal U=\operatorname{span}\{\|\Phi\|^2,\Phi_1,\ldots,\Phi_D\}$, hence $D+1\ge D_\star$. Equality gives $\mathcal U=\mathcal V$, making every $\Phi_i$ and $\|\Phi\|^2$ polynomials of degree at most two on the open set. If $q_i$ is the homogeneous quadratic part of $\Phi_i$, the quartic part of $\|\Phi\|^2$ is $\sum_iq_i^2=0$. Thus all $q_i$ vanish, every $\Phi_i$ is affine, and $\dim\mathcal U\le d+1<D_\star$, a contradiction.
\eop

\noindent\underline{\bf Issues:}
The issue is the substantial dimensionality increase introduced by quadratic lifting. The dimension-optimal exact lifting still has $\Theta(d^2)$ dimensions. Thus, the lifted vectors require $\Theta(Nd^2)$ storage. More importantly, applying a conventional point ANN index in this lifted space substantially increases both index construction and query costs, since distance evaluations and graph construction operate over the much higher-dimensional representations. {As shown in Exps.~1 and 2,} this dimensionality expansion leads to significantly higher construction overhead and substantially lower query efficiency than searching in the original space.

\section{Line-Enhanced Proximity Graph}
\label{sec:overview}

Section~\ref{sec:limit} identifies three limitations. Enumerating all point pairs to construct line-specific structures is expensive, and quadratic lifting increases the dimension of every distance computation. Moreover, reusing a proximity graph for LANN search is difficult, since points close to one line may occupy several regions, whereas the graph mainly connects nearby points. Thus, a beam search can explore one region and stop before reaching another.
To address these issues, we design a line-enhanced proximity graph, which reuses the original proximity graph and enhances it with a limited number of connections between regions.

\stitle{Connect Near-Line Regions:}
\label{sec:pair-idea}
Our key idea is to use original edges in the proximity graph for local exploration and add \textbf{pair edges} to reach other near-line regions.
Within a small region, points are close to each other, and the proximity graph already provides useful local connections.
The missing connection is often between two regions far apart along the line.
When two regions are close to a query line but far apart, we can find a line linking them with a direction highly similar to the query line.
To see this, let distinct $p,q\in S_r(L)$ and let $u$ be the query direction. Since $q-p=(\tau_L(q)-\tau_L(p))u+e_L(q)-e_L(p)$, its perpendicular component has norm at most $\norm{e_L(q)}+\norm{e_L(p)}\le2r$. Thus, $\sin\angle(q-p,u)\le\min\{1,2r/\norm{q-p}\}.$
When the two points are far apart relative to $r$, i.e., $\norm{q-p}\gg r$, their secant is nearly parallel to $L$. For example, in Fig.~\ref{fig:motivation-linePG}, the line between $p_9$ and $p_8$ has a direction highly similar to $L$.

We therefore consider using these edges between two near-line regions, named pair edges, to provide an alternative route.
On the original proximity graph, reaching a different near-line region may require intermediate points that are farther from the line and are rejected by the beam search.
A pair edge provides an alternative to this detour; local exploration resumes through ordinary edges at its target.
For example, reconsider Example~\ref{ex:direct} and Fig.~\ref{fig:motivation-linePG}. Adding a pair edge $p_9\to p_8$ gives the search from $p_9$ access to $R_2=\{p_4,p_7,p_8\}$, and expanding $p_8$ finds $p_4$ and $p_7$.
To keep the line-enhanced proximity graph within the degree limit $M$, adding pair edges may require pruning some original edges in the proximity graph, e.g., we substitute the edge $p_9 \to p_{10}$ with the edge $p_9 \to p_8$ in Fig.~\ref{fig:motivation-linePG} to maintain the degree limit of 3.

Let $G=(V,E)$ be the original proximity graph. A {pair edge} is an added edge $x\to y$ between points in two selected regions. Let $\Epair$ be these edges and $\Ereg\subseteq E$ be the retained original edges. The \textbf{line-enhanced proximity graph} is $G^+=(V,\Ereg\cup\Epair)$.

However, to construct a line-enhanced proximity graph, there are $\Theta(N^2)$ potential point pairs to form pair edges. Evaluating them individually is costly; we thus group points into regions and first select potential {region pairs} via the hierarchy and quotient graph.

\begin{figure}[t]
  \centering

  \begin{minipage}[t]{0.42\linewidth}
    \vspace{0pt}
    \centering
    \includegraphics[width=0.87\linewidth]{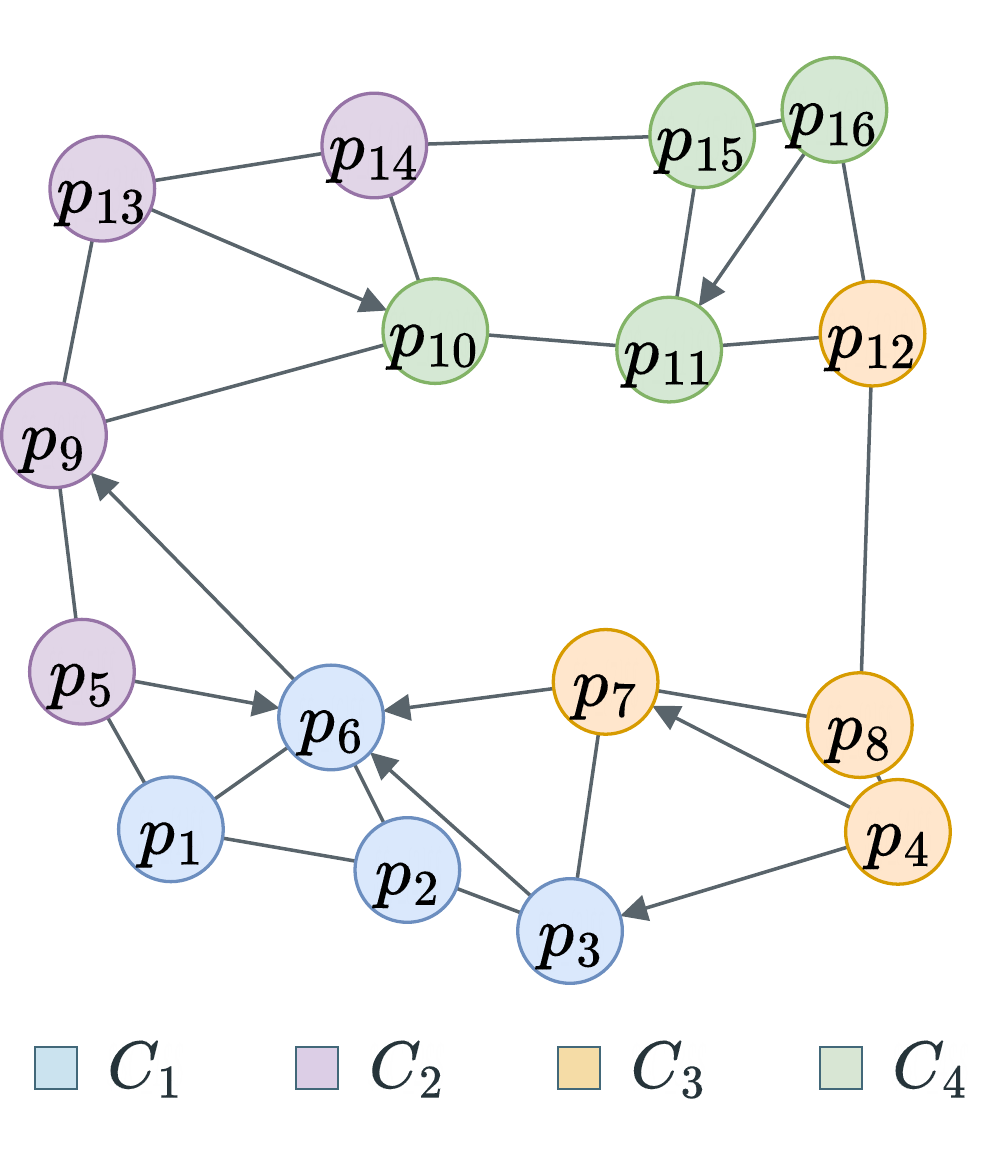}
    \vspace*{-3mm}
    \caption{Region partition of the example in Fig.~\ref{fig:motivation-linePG}}
    \label{fig:example-region}
  \end{minipage}
  \hfill
  \begin{minipage}[t]{0.55\linewidth}
    \vspace{0pt}
    \centering
    \begin{minipage}[t]{\linewidth}
      \vspace{0pt}
      \centering
      \includegraphics[width=0.8\linewidth]{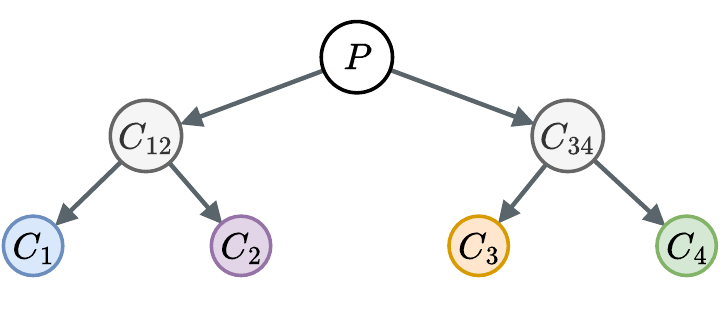}
      \vspace*{-3mm}
      \caption{Hierarchy of Fig.~\ref{fig:example-region}}
      \label{fig:example-hierarchy}
    \end{minipage}

    \vspace{1mm}

   \begin{minipage}[t]{0.47\linewidth}
      \vspace{0pt}
      \centering
      \includegraphics[width=0.52\linewidth]{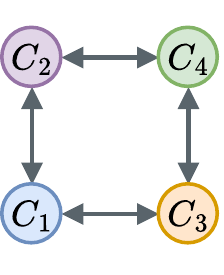}
      \vspace*{-3mm}
      \caption{Quotient graph}
      \label{fig:example-quotient}
    \end{minipage}
    \hfill
    \hspace{-2mm}
    \begin{minipage}[t]{0.47\linewidth}
      \vspace{0pt}
      \centering
      \includegraphics[width=0.52\linewidth]{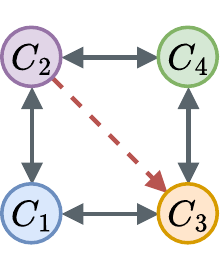}
      \vspace*{-3mm}
      \caption{Candidate arc in Fig.~\ref{fig:example-quotient}}
      \label{fig:example-quotient-arc}
    \end{minipage}
  \end{minipage}
  \vspace{-4mm}
\end{figure}

\stitle{Region Pairs on Hierarchy:}
We construct a balanced hierarchy using both the vectors and the original proximity graph. Vector geometry groups points at a local spatial scale. Graph smoothing adjusts the split values so that neighboring points influence each other's assignment.
{Figs.~\ref{fig:example-region} and \ref{fig:example-hierarchy} show the complete hierarchy for the example in Fig.~\ref{fig:motivation-linePG}.} The root first splits into $C_{12}=C_1\cup C_2$ and $C_{34}=C_3\cup C_4$. Each has eight points and is split once more. The left child gives $C_1=\{p_1,p_2,p_3,p_6\}$ and $C_2=\{p_5,p_9,p_{13},p_{14}\}$, and the right child gives $C_3=\{p_4,p_7,p_8,p_{12}\}$ and $C_4=\{p_{10},p_{11},p_{15},p_{16}\}$.
Each internal node in the hierarchy represents a region of points which is the disjoint union of its two children. Leaves partition the whole dataset, so every point $p$ has exactly one owner $\ell(p)$, e.g., $\ell(p_1)=C_1$ and $\ell(p_5)=C_2$.

As discussed before, we aim to find pair edges that link regions close to a line yet far apart, and far-apart regions close to the same line have a common approximate direction.
Different subtrees provide candidate spatial alternatives, but do not by themselves guarantee separation or proximity to a common line.
Therefore, we use the hierarchy to first propose partners from different sibling subtrees, e.g., for node $C_2$ in the hierarchy, its sibling is $C_1$, and its parent's sibling is $C_{34}$, containing $C_3,C_4$. 
A score then compares separation against local spread and support near the representative secant. The details are given in Section~\ref{sec:score-design}.

\stitle{Region Pairs on Quotient Graph:}
Based on the hierarchy, we then construct a {quotient graph}: each leaf becomes a vertex, and an arc records an original edge between two different leaves. Thus, the quotient graph exposes how the original graph connects those regions.
{For example, Fig.~\ref{fig:example-quotient} shows the four leaves from Fig.~\ref{fig:example-hierarchy}.} Its arcs are $C_1\leftrightarrow C_2$, $C_1\leftrightarrow C_3$, $C_2\leftrightarrow C_4$, and $C_3\leftrightarrow C_4$. Each double arrow stands for two directed arcs, e.g., the four edges from $C_2$ to $C_4$ include $p_9\to p_{10}$ and $p_{14}\to p_{15}$; they give one $C_2\to C_4$ arc.
We next use quotient graph topology to propose region pairs without a short directed route between them. Such pairs can expose detours worth shortening. {For instance, Fig.~\ref{fig:example-quotient-arc} illustrates how a $C_2\to C_3$ connection can shorten an inter-region detour.}
However, the existence or length of a quotient route says nothing by itself about its point-to-line distances. The same score, based on geometric separation and local spread and support near the representative secant, is thus evaluated to identify the region pairs.

Therefore, for selecting candidate region pairs, the hierarchy exposes spatial alternatives at several scales, and the quotient graph exposes how the original graph connects those regions. {Sections~\ref{sec:cluster} and \ref{sec:candidates} give the construction and candidate-generation details.}

\stitle{Pair Edge Selection:}
A retained candidate region pair $(A,B)$ must next become an actual edge $p\to y$, for $p\in A$ and $y\in B$. Using only the centroid $c_B$ of $B$ for every node $p\in A$ could concentrate entrances at one point and miss useful endpoint alternatives. We therefore form a small {representative bank} $\mathcal{R}(B)$ for each leaf: points nearest $c_B$, followed by a bounded number of farthest distinct points. Then, for each point $p\in A$, we consider selecting edges from $p$ to nodes $y\in B$ based on the following rules.

\textbf{(1)} The selected regions $A$ and $B$ are separated; as discussed above, next, they should be close to a line, i.e., a useful pair edge should follow a direction that is supported by multiple points in the target region. Hence, for a candidate line $p\to y$ where $y\in \mathcal{R}(B)$, we use $J(p,y;B)$ to evaluate if other points in $B$ support the line. Here $J(p,y;B)=\sum_{z\in B\setminus \{y\}}{1}/({1+\psi_{p,y}(z)})$ where $\psi_{p,y}(z)={\dist^2(z,p+\R(y-p))}/{\norm{y-p}^2}$ is the normalized squared distance from $z$ to the line through $p$ and $y$.
A larger $J(p,y;B)$ indicates that the direction $p\to y$ is supported by more points in the target region.
Among the candidate endpoints examined for the same target region $B$, we prefer the valid endpoint with the largest $J(p,y;B)$. An endpoint is valid only if $\norm{y-p}>0$ and its edge belongs to neither the original edge set $E$ nor the current $\Epair$.
For example, reconsider Fig.~\ref{fig:example-region}, where $p_9=(-0.28,1.84)$ and $C_3=\{p_4=(3.04,0.28),p_7=(1.89,0.87),p_8=(2.89,0.70),p_{12}=(2.94,2.24)\}$. Assume $\mathcal R(C_3)$ here is $C_3$. We have $J(p_9,p_4;C_3)=2.81$, $J(p_9,p_7;C_3)=2.66$, $J(p_9,p_8;C_3)=2.83$ and $J(p_9,p_{12};C_3)=2.43$. Thus $p_8$ is selected, since it receives the largest support, as shown in Fig.~\ref{fig:motivation-linePG}.

\textbf{(2)} After choosing an endpoint for each candidate target region, we further discourage pair edges whose directions are redundant with either previously selected pair edges or the original proximity graph.
Let $\chi(p;y,z)$ denote the squared cosine similarity between $y-p$ and $z-p$, with $\chi=0$ for a degenerate direction.
(i) For selected pair edges, use $\nu(p,y)=1-\max_{z\in F_p\setminus\{y\}}\chi(p;y,z)$, where $F_p$ contains the pair endpoints already selected for $p$.
A large $\nu(p,y)$ therefore indicates that the candidate $p\to y$ introduces a direction different from those already selected for $p$.
(ii) For the original proximity graph, use $\alpha(p,y)=1-\max_{z\in N_G^+(p,y)}\chi(p;y,z)$, where $N_G^+(p,y)=\{z\in N_G(p):(y-p)^\top(z-p)>0\}$ contains the existing neighbors of $p$ in forward half-space of $y-p$.
Thus, a large value means the edge direction is poorly represented by existing edges.

\begin{algorithm}[t]
\small  
    \SetVline     \SetFuncSty{textsf}
    \SetArgSty{textsf}
 \caption{\kw{BuildLPG}($\Pset,G,\Mpair$)}
 \label{algo:construct-linePG}
 \Input{dataset $\Pset$, proximity graph $G=(V,E)$ with degree limit $M$, and pair edge limit $\Mpair$}
 \Output{line-enhanced proximity graph $G^+$}

{$\mathcal{T}\leftarrow$ the hierarchy built based on $\Pset$ and $G$};\hfill\tcp{Section~\ref{sec:smoothing}}
\State{$Q\leftarrow$ the quotient graph of $\mathcal{T}$, where $V(Q)=\{\text{leaf nodes in }\mathcal{T}\}$ and $E(Q)=\{(u,v):u\ne v,\ \exists(p,q)\in E,\ \ell(p)=u,\ell(q)=v\}$}
$\mathcal{C}\leftarrow$\kw{SelectRegions}($\mathcal T,Q$);\hfill\tcp{Algorithm~\ref{algo:select-regions}}
\State{$\Ereg\leftarrow E$; $\Epair\leftarrow \emptyset$}
\For{each $(A,B)\in \mathcal{C}$}{
    \For{each $p\in A$}{
        \State{$y^*\leftarrow \arg\max_{y\in \mathcal{R}(B)} J(p,y;B)$}
        \State{$\Epair\leftarrow \Epair\cup\{(p,y^*)\}$}
            \If{the edge count from $p$ in $\Epair$ exceeds $\Mpair$}
            {
                \State{Remove $(p,y')$ with minimum $\nu(p,y')\alpha(p,y')$ from $\Epair$}
            }
            \If{the edge count from $p$ in $\Epair\cup\Ereg$ exceeds $M$}{
                \State{Remove $(p,y')$ with maximum $\delta(p,y')$ from $\Ereg$}
            }
    }
}
\Return{$G^+\leftarrow(V,\Ereg\cup\Epair)$}
\end{algorithm}

Algorithm~\ref{algo:construct-linePG} gives the construction of line-enhanced proximity graph $G^{+}$.
Given a proximity graph with degree limit $M$ built for point ANN search and a pair-edge limit $\Mpair$, we first build the hierarchy and quotient graph to select candidate region pairs (lines 1-3). For each point in a source region (lines 5-6), we select a target endpoint maximizing $J(p,y;B)$ from the representative bank of the target region (line 7) and insert it into pair edge set (line 8).
When the newly selected pair edge leads to exceeding the degree limit, we remove the edge based on direction diversity scores $\nu(p,y)$ and $\alpha(p,y)$ if the pair-edge count exceeds $\Mpair$ (lines 9-10), or prune the longest original edge otherwise (line 11-12). Note that the edges selected in proximity graph $G$ are already pruned by RNG pruning strategy~\cite{hnsw}, hence, we only need to prune the longest distance edge to satisfy the degree limit $M-\Mpair$ here. The resulting graph is $G^+=(V,\Ereg\cup\Epair)$ (line 13).

\section{Hierarchy and Quotient Graph Constructions}
\label{sec:cluster}

In this section, we give the construction details for the hierarchy $\mathcal T$ and quotient graph $Q$ introduced in Section~\ref{sec:overview}.

\label{sec:smoothing}
The hierarchy is built by recursive balanced partitioning. Before each split, we smooth a scalar split signal derived from the vectors over the original graph, so that the assignment reflects both point locations and local graph connections.

\stitle{Partition via Graph Smoothing:}
For a hierarchy node containing point set $C$, let $c_C=\frac{1}{|C|}\sum_{p\in C}p$ be its centroid, and choose the coordinate of maximum variance, $ j=\arg\max_{\ell\in[d]}\frac{1}{|C|}\sum_{p\in C}(p_\ell-c_{C,\ell})^2.$
The initial split value of $p\in C$ is $ f_p^{(0)}=p_j-c_{C,j}.$
Partitioning these values at the median gives a balanced vector-based geometric split. Intuitively, this cuts region $C$ at its median along the coordinate axis of greatest spread.

To account for the proximity graph, we consider the subgraph induced by $C$. Specifically, neighbors in $N_G(p)\cap C$ contribute to the split value of $p$. For an edge $p\to q$ in this subgraph, define $ w_{pq}=\frac{1}{1+\norm{p-q}^2/\sigma_C^2}$, where $\sigma_C^2=\frac{1}{|C|}\sum_{p\in C}\norm{p-c_C}^2$ is the mean squared distance to the centroid. Shorter edges therefore receive larger weights. For each $p$ with a nonempty induced neighbor list, normalize over $q\in N_G(p)\cap C$: $W_{pq}=\frac{w_{pq}}{\sum_{q'\in N_G(p)\cap C}w_{pq'}},$ and set all other entries to zero. If $\sigma_C^2=0$, use equal weights on the existing induced edges. If $N_G(p)\cap C=\emptyset$, set $W_{pp}=1$. These conventions keep every row sum equal to one.

At each round $t$, update $ f^{(t+1)}=\tfrac12\bigl(f^{(t)}+Wf^{(t)}\bigr).$
The updates use the values from the preceding round. They change only the split signal, not the point coordinates or graph edges. The following theorem formalizes how smoothing encourages neighboring points to have similar split values.

\begin{theorem}\label{thm:smoothing}
Let $W$ be the row-stochastic matrix above, and let $\pi$ be any stationary distribution of $W$, i.e., $\pi_p\ge0$, $\sum_p\pi_p=1$, and $\pi^\top W=\pi^\top$. The stationary mean $\bar f=\sum_p\pi_p f_p^{(0)}$ is preserved. For $ V_t=\sum_{p\in C}\pi_p(f_p^{(t)}-\bar f)^2,$ each smoothing round satisfies $ V_t-V_{t+1}\ge\frac14\sum_{p,q\in C}\pi_pW_{pq}
 \bigl(f_p^{(t)}-f_q^{(t)}\bigr)^2\ge0.$
Moreover, every new split value lies between the minimum and maximum values of preceding round.
\end{theorem}

\proofsketch
\sloppy
Stationarity gives
$\pi^\top f^{(t+1)}=\pi^\top f^{(t)}$.
Put $g=f^{(t)}-\bar f\mathbf1$, and write
$\langle g,h\rangle_\pi=\sum_p\pi_pg_ph_p$ and
$\|g\|_\pi^2=\langle g,g\rangle_\pi$.
By Jensen's inequality~\cite{Boyd_Vandenberghe_2004}, the unit row sums of $W$, and stationarity, $\|Wg\|_\pi^2
 =\sum_p\pi_p\left(\sum_qW_{pq}g_q\right)^2
 \le\sum_{p,q}\pi_pW_{pq}g_q^2
  =\sum_q\pi_qg_q^2=\|g\|_\pi^2.$
Also, expanding the weighted edge differences gives $\sum_{p,q}\pi_pW_{pq}(g_p-g_q)^2
 =2\|g\|_\pi^2-2\langle g,Wg\rangle_\pi.$
 Combining these identities with the update yields $V_t-V_{t+1}=\tfrac14\sum_{p,q}\pi_pW_{pq}(g_p-g_q)^2+\tfrac14\bigl(\|g\|_\pi^2-\|Wg\|_\pi^2\bigr)\ge\tfrac14\sum_{p,q}\pi_pW_{pq}(g_p-g_q)^2.$
Subtracting a common mean does not change pairwise differences, giving the stated bound. Finally, each new value is a convex combination of the preceding values, so it lies within their range. 
\eop

Iterating smoothing to convergence can suppress the geometric information in the initial split signal. In particular, when the induced graph is strongly connected, the values converge to the same stationary mean, losing their geometric separation. We thus use a small number $t_s$ of rounds, with $t_s=4$ by default in Section~\ref{sec:exp}.

\begin{figure}
  \subfigure[Vector-based median split]
  {
      \includegraphics[width=0.4\columnwidth]{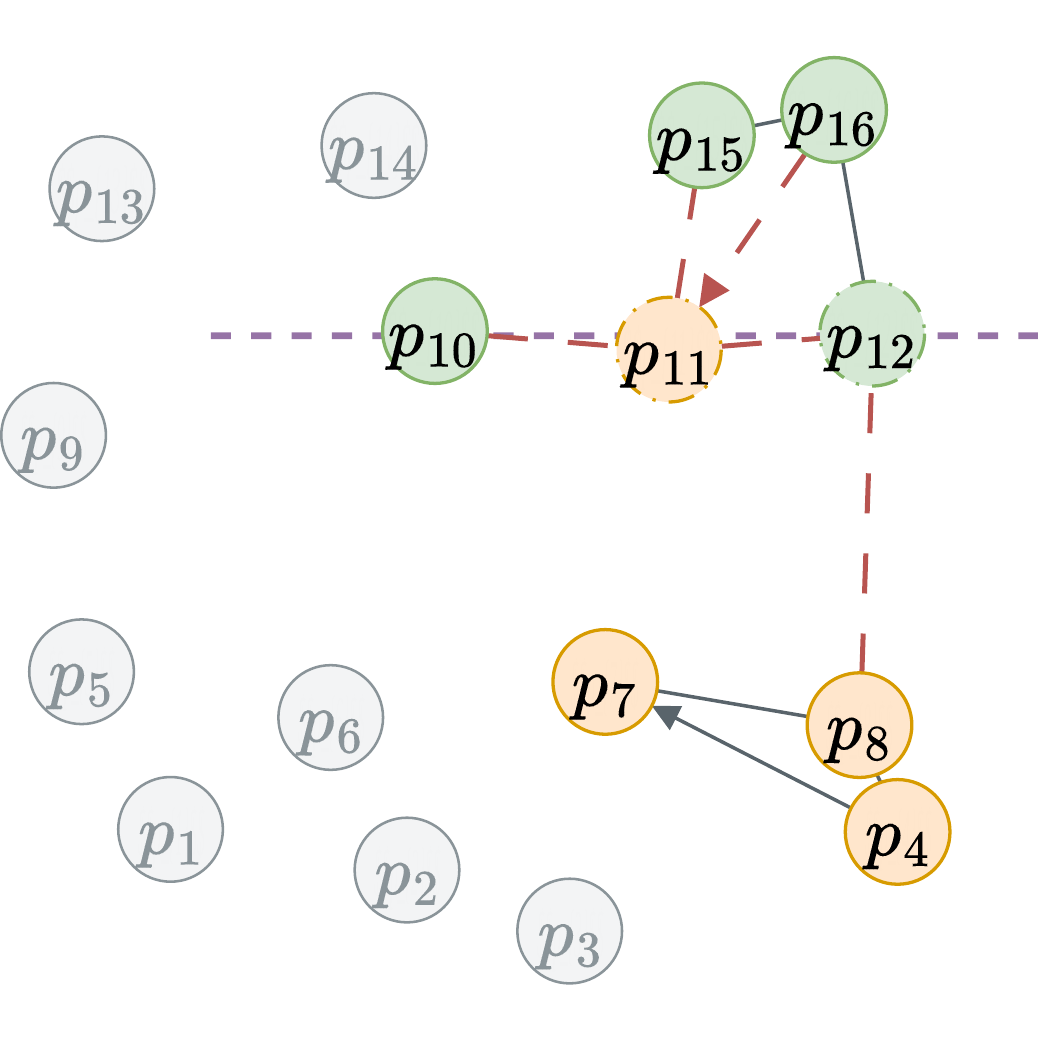}
      \label{fig:partition-raw}
  }
\hspace{4mm}
  \subfigure[Partition after graph smoothing]
  {
      \includegraphics[width=0.4\columnwidth]{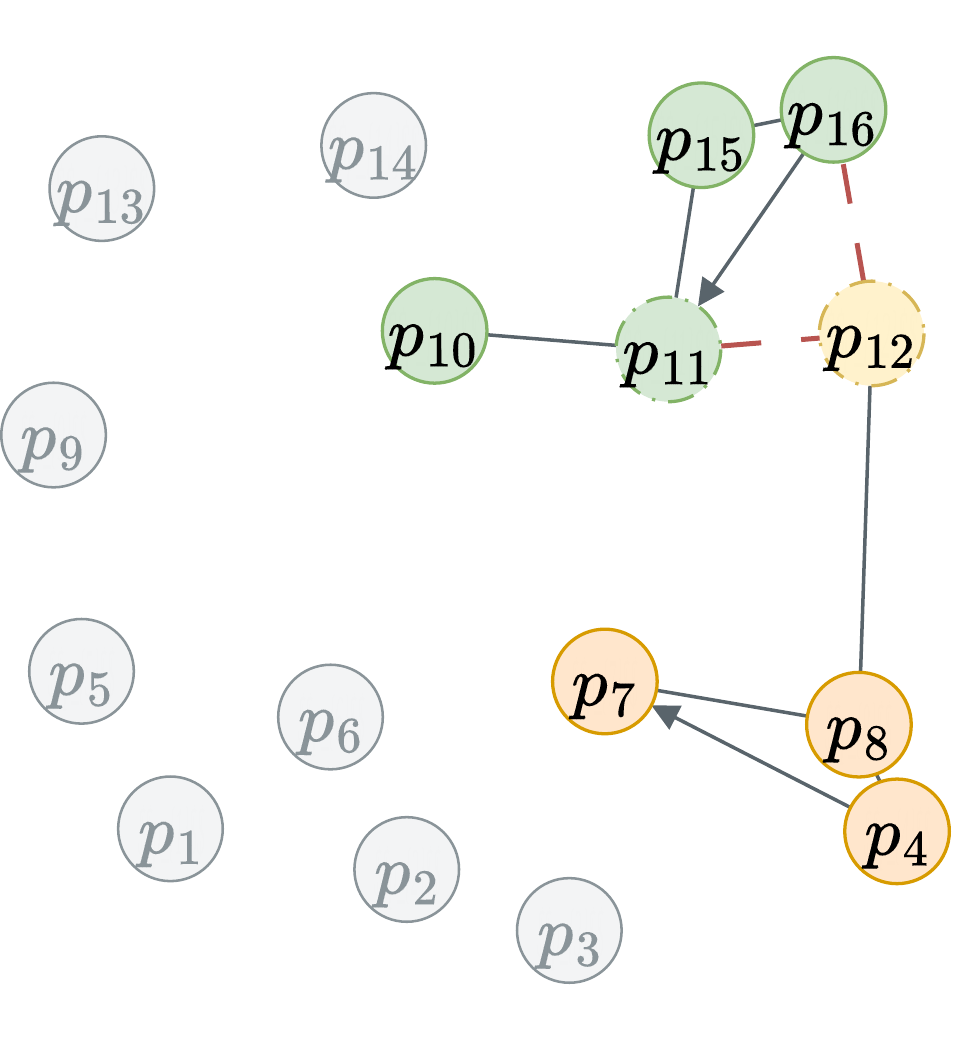}
        \label{fig:partition-smoothed}
  }

\vspace{-4mm}
\caption{An example of region partition}
\label{fig:smoothing}
\vspace{-3mm}
\end{figure}

\begin{example}
\label{ex:smoothing}
Reconsider the right subtree $C_{34}$ in Fig.~\ref{fig:example-hierarchy}. Fig.~\ref{fig:smoothing} compares two splits of this same parent. Its maximum-variance coordinate is $y$. A raw median split, i.e., split via the horizontal line (the dashed purple line in Fig.~\ref{fig:partition-raw}), places $p_{11}$ in the lower child and $p_{12}$ in the upper child, since $p_{11}$'s $y$ coordinate is smaller than $p_{12}$'s. After smoothing, their assignments are exchanged. Thus $p_{12}$ is grouped with its neighbor $p_8$, while $p_{11}$ is grouped with $p_{10}$ and $p_{15}$. The number of directed edges crossing this split decreases from nine to four, as highlighted in dashed red lines.
\end{example}

\stitle{Hierarchy Construction:}
After smoothing, order points in $C$ by $f_p^{(t_s)}$, and split at the median. The two children contain $\lfloor|C|/2\rfloor$ and $\lceil|C|/2\rceil$ points. Recursive partitioning starts at $C=\Pset$ and continues until every leaf contains at most $b$ points. We set $b=32$ by default in Section~\ref{sec:exp}, whereas the example shown in Fig.~\ref{fig:example-hierarchy} uses $b=4$.

\stitle{Quotient Graph Construction:}
\label{sec:quotient}
Once the hierarchy is built, we can obtain the leaf-owner record $\ell$ for each point in $\Pset$. Next, we can construct the quotient graph by scanning the original edges. Specifically, for each $(p,q)\in E$, add the arc $(\ell(p),\ell(q))$ when $\ell(p)\ne\ell(q)$, then remove duplicate arcs. Multiple point edges therefore give only one quotient arc. This produces Fig.~\ref{fig:example-quotient} without rebuilding a proximity graph for any leaf. Note that the averaging weights in $W$ are not used as quotient-arc weights: $Q$ records only the original inter-leaf connections.

We next analyze the cost of constructing the hierarchy and quotient graph.
Let $m=|E|=O(MN)$ and $h=O(1+\log\max\{1,N/b\})$ be the number of hierarchy levels.
Using linear-time median selection for the balanced split~\cite{BlumFPRT73}, the hierarchy construction takes $ O\left(h\bigl(Nd+md+t_s(N+m)\bigr)\right)$ time.
Computing centroids, coordinate variances, and mean squared spreads costs $O(Nd)$ per level. Using sparse neighbor rows, forming and normalizing the geometric edge weights costs at most $O(md+N)$; these weights stay fixed during the $t_s$ rounds at that node.
Each round costs $O(N+m)$ per level, and median partitioning costs $O(N)$.
Quotient construction takes expected $O(N+m)$ time using hash-based deduplication. Hence, the total cost of constructing the hierarchy and quotient graph is $O\left(h\bigl(Nd+md+t_s(N+m)\bigr)\right)$.

\section{Candidate Region Pair Selection}
\label{sec:edges}

Section~\ref{sec:overview} gives the endpoint score $J$, the directional factors $\nu,\alpha$, and the edge-replacement procedure. What remains is to identify the region pairs supplied to that procedure. To generate candidate region pairs, we first generate candidates from $\mathcal T$ and $Q$, then retain those with high geometric separation and secant support. The selected region pairs are finally converted into point-level edges.

\subsection{Candidate Region Pairs}
\label{sec:candidates}

As discussed in Section~\ref{sec:overview}, candidate region pairs are generated from the hierarchy $\mathcal T$ and quotient graph $Q$, which both return ordered pairs of distinct leaves.

\stitle{Pairs from Sibling Subtrees on Hierarchy:}
We use the internal hierarchy nodes to organize the search for partners, and use $s_h$ to limit the number of leaves sampled from each sibling subtree. Specifically, as shown in Algorithm~\ref{algo:select-regions}, for a source leaf $A$, start at $v=A$ and move toward the root (lines~\ref{ln:select-init}--\ref{ln:select-parent}). At each nonroot node $v$, let $u$ be its sibling and $\mathcal L(u)$ its descendant leaves. Select $\min\{s_h,|\mathcal L(u)|\}$ leaves uniformly without replacement from $\mathcal L(u)$ and add them as candidate target regions of $A$. Thus, a small sibling subtree contributes all its leaves; a large one contributes only a sample. The same per-subtree budget prevents the many leaves at a coarse level from exhausting the candidate generation work. Different sibling subtrees along this path are disjoint and together contain all leaves other than $A$.
To perform this sampling without enumerating a large subtree on every visit, a depth-first traversal first numbers the leaves and records the contiguous interval of leaf ranks of each node (line~\ref{ln:select-index}). Then, at line~\ref{ln:select-sample}, we can draw distinct ranks uniformly from that interval and return the corresponding leaves.
For example,
reconsider source $C_2$ in Fig.~\ref{fig:example-hierarchy} with $s_h=1$. Its sibling is $C_1$, so the first step adds $C_1$. After moving to $C_{12}$, the sibling is $C_{34}$, whose leaf interval contains $C_3,C_4$. It then chooses either $C_3$ or $C_4$, each with probability $1/2$.

\stitle{Pairs Beyond Short Quotient Paths:}
A useful candidate region pair should not merely repeat a connection available through a few region transitions. Let $\operatorname{hop}_Q(A,B)$ be the shortest directed hop distance from $A$ to $B$, taking value $\infty$ when no path exists. For a positive hop radius $\kappa$ and leaf set $\mathcal L$, define $ \mathcal N_Q^\kappa(A)=\{B\in\mathcal L:\operatorname{hop}_Q(A,B)\le\kappa\}, \mathcal F_\kappa(A)=\mathcal L\setminus\mathcal N_Q^\kappa(A).$
Only leaves in $\mathcal F_\kappa(A)$ are eligible for the quotient source.
This criterion has a direct connection to the original proximity graph. Any path in $G$ from $p\in A$ to $y\in B$ induces a walk in $Q$ with no more edges, after consecutive repetitions of leaf identifiers are removed. Thus, $B\in\mathcal F_\kappa(A)$ rules out every original point-graph path of at most $\kappa$ edges between the two leaves. An added edge $p\to y$ provides a one-step alternative.
We use the same parameter $s_h$ to bound the candidate region pairs for each leaf node $A\in \mathcal L$. Each region $A$ will generate at most $n_w=s_h\lceil\log_2|\mathcal L|\rceil$ region pairs, such that the number of region pairs from the hierarchy and the quotient graph have the same scale.
Due to the small-world diameter in proximity graphs~\cite{hnsw,WangXY021,milgram1967small,MalkovPLK14}, $\kappa$ does not have to be large, we set $\kappa=3$ by default. 

Exactly obtaining the set $\mathcal F_\kappa(A)$ for each leaf $A\in \mathcal L$ is costly, although only a few targets in $\mathcal F_\kappa(A)$ are needed. Therefore, as shown in Algorithm~\ref{algo:select-regions}, we reverse this order: instead of first obtaining $\mathcal F_\kappa(A)$ and then sampling from it, we first draw one shared pool $D=(D_0,\ldots,D_{n_p-1})$ of $n_p$ distinct leaves uniformly from $\mathcal L$ (line~\ref{ln:select-proposals}) and then determine only which members of $D$ are reachable within $\kappa$ hops.
We use $n_p=\min\{|\mathcal L|,128n_w\}$ by default to provide sufficient candidate regions for each leaf node.
We use one bit per node in $D$, where bit $j$ of $X_D^{(t)}(A)$ records whether $D_j$ is reachable from $A$ in at most $t$ hops. Initialize it to one exactly when $A=D_j$ (line~\ref{ln:select-seedbits}), and perform $\kappa$ synchronous updates: $X_D^{(t+1)}(A)=X_D^{(t)}(A)\;\lor\!\bigvee_{Z\in N_Q(A)}X_D^{(t)}(Z)$ (lines~\ref{ln:select-rounds}-\ref{ln:select-topology}).
A row reads its outgoing neighbors' previous masks, preserving the direction $A\to D_j$; two buffers prevent a newly written value from propagating twice in one round. The zero bits at valid target positions of the final mask identify $U_A=D\cap\mathcal F_\kappa(A)$ (line~\ref{ln:select-eligible}), and then we can independently select a uniform subset $\mathcal W_A$ of size $\min\{n_w,|U_A|\}$ from $U_A$ as the candidate target regions for $A$ (line~\ref{ln:select-quotient-sample}).
This uses the bit-parallel traversal principle in~\cite{ThenKCHPK0V14}, but applies it only to sampled targets.
The following theorem proves that sampling targets before checking short-hop reachability does not favor any eligible leaf when the randomness of $D$ is included.

\begin{theorem}\label{thm:sampled-quotient}
Fix a source $A$ with $\mathcal F_\kappa(A)\ne\emptyset$. For every possible returned size $s$ and every $B\in\mathcal F_\kappa(A)$, $ \Pr\!\left[B\in\mathcal W_A\mid |\mathcal W_A|=s\right]
 ={s}/{|\mathcal F_\kappa(A)|},$
where the probability includes both the random pool $D$ and the subsequent subsampling, and the conditioned event has positive probability.
\end{theorem}

\proofsketch
Fix $A$ and write $F=\mathcal F_\kappa(A)$ and $m=|F|>0$.
For any two leaves $B,C\in F$, consider the permutation that exchanges their identifiers and fixes every other leaf. Since $D$ is uniform among subsets of the prescribed size, applying this permutation leaves its distribution unchanged. It also preserves $|D\cap F|$, hence the returned size $\min\{n_w,|D\cap F|\}$.
Conditional on any pool, the subsequent subsampling is uniform within $D\cap F$. Exchanging $B,C$ thus exchanges their inclusion events without changing their probabilities or the event $|\mathcal W_A|=s$. For every possible $s$, all $m$ eligible leaves consequently have the same conditional inclusion probability; denote it by $\rho_s$. As $\mathcal W_A\subseteq F$, $s=\mathbb E\bigl[|\mathcal W_A|\mid |\mathcal W_A|=s\bigr]=\sum_{B\in F}\Pr\bigl[B\in\mathcal W_A\mid |\mathcal W_A|=s\bigr] =m\rho_s.$
Thus $\rho_s=s/m$. The probability averages over both random stages; it is not a uniformity claim after fixing $D$, nor after the later score-based top-$r$ selection. 
Sharing $D$ can also correlate the samples of different sources.
\eop

\begin{algorithm}[t]
\small
\SetVline
\SetFuncSty{textsf}
\SetArgSty{textsf}
\caption{\kw{SelectRegions}($\mathcal T,Q$)}
\label{algo:select-regions}
\Input{hierarchy $\mathcal T$, quotient graph $Q$}
\Parameter{budgets $s_h$, hop radius $\kappa$, and region pair limit $r$}
\Output{ordered list $\mathcal C$ of selected region pairs}
Number leaves by DFS; store each subtree's leaf interval\;\nllabel{ln:select-index}
$n_w\leftarrow s_h\lceil\log_2|V(Q)|\rceil$; $n_p\leftarrow\min\{128n_w,|V(Q)|\}$\;
$D\leftarrow$ a uniform $n_p$-subset of $V(Q)$, in random order\;\nllabel{ln:select-proposals}
$X_D(Z)\leftarrow$ a zero $n_p$-bit mask for each leaf $Z$\;\nllabel{ln:select-masks}
Set bit $j$ of $X_D(D_j)$ to one for each $0\le j<n_p$\;\nllabel{ln:select-seedbits}
\For{$t\leftarrow1$ \KwTo $\kappa$\nllabel{ln:select-rounds}}{
 \For{each leaf $Z$ in $Q$\nllabel{ln:select-propagate-source}}{
  $Y_D(Z)\leftarrow X_D(Z)$\;\nllabel{ln:select-copy}
  \textbf{for} {each $Z'\in N_Q(Z)$\nllabel{ln:select-propagate-neighbor}} \textbf{do}  {
   $Y_D(Z)\leftarrow Y_D(Z)\lor X_D(Z')$\;\nllabel{ln:select-propagate}
  }
 }
 Swap mask arrays $X_D$ and $Y_D$\;\nllabel{ln:select-topology}
}
$\mathcal C\leftarrow\emptyset$\;\nllabel{ln:select-empty}
\For{each leaf $A$ in identifier order\nllabel{ln:select-source}}{
 $U_A\leftarrow\{D_j\in D:\text{bit }j\text{ of }X_D(A)=0\}$\;\nllabel{ln:select-eligible}
 $\mathcal W_A\leftarrow$ a uniform $\min\{n_w,|U_A|\}$-subset of $U_A$\;\nllabel{ln:select-quotient-sample}
 $\mathcal B_A\leftarrow\mathcal W_A$; $v\leftarrow A$\;\nllabel{ln:select-init}
 \While{$v$ is not the root\nllabel{ln:select-up}}{
  $u\leftarrow$ the sibling of $v$\;\nllabel{ln:select-sibling}
  Sample $\min\{s_h,|\mathcal L(u)|\}$ leaves in the subtree rooted at $u$ to insert into $\mathcal B_A$\;\nllabel{ln:select-sample}
  $v\leftarrow$ the parent of $v$\;\nllabel{ln:select-parent}
 }
 Deduplicate $\mathcal B_A$; remove $A$\; \nllabel{ln:select-cap}
 Append up to $r$ pairs $(A,B)$ to $\mathcal C$ in descending $S(A,B)$\;\nllabel{ln:select-keep}
}
\Return{$\mathcal C$}\nllabel{ln:select-return}
\end{algorithm}

\stitle{Merging the Candidate Region Pairs:}
As shown in Algorithm~\ref{algo:select-regions} and discussed above, each leaf $A$ generates a candidate set $\mathcal B_A$ from the hierarchy (line~\ref{ln:select-sample}) and the quotient graph (line~\ref{ln:select-quotient-sample}).
The two sources express different preferences: the hierarchy offers spatial alternatives at several scales, while the quotient sample offers destinations without short directed routes. We deduplicate the union of these two sets (line~\ref{ln:select-cap}), and each candidate will be scored (line~\ref{ln:select-keep}), as described in Section~\ref{sec:score-design}. The best $r$ region pairs will be returned to generate pair edges (lines~\ref{ln:select-keep}-\ref{ln:select-return}).

\subsection{Scoring Candidate Regions}
\label{sec:score-design}

After obtaining the candidate region pairs, we consider scoring them based on geometric considerations from Section~\ref{sec:overview}: a pair edge should span more than the local extent of its regions, and this connection should describe a direction supported by the regions' points. We use $g(A,B)$ for the first consideration and $h(A,B)$ for the second.
For each final leaf $A$, compute its centroid $c_A$ and mean squared spread $\sigma_A^2$ from the original vectors, as in Section~\ref{sec:smoothing}. The representative secant of $(A,B)$ is the infinite line through the two centroids: $\lambda_{AB}=\{c_A+t(c_B-c_A):t\in\R\}$. The Euclidean distance between these centroids is $d_{AB}=\norm{c_B-c_A}$, which measures the separation of the regions' representative locations.

\stitle{Separation Relative to Local Spread:}
A large centroid distance alone is insufficient, since two large regions may have distant centroids without being well separated at their own scale. We thus compare $d_{AB}^2$ with $\sigma_A^2+\sigma_B^2$, the combined mean squared spread. Define $g(A,B)={d_{AB}^2}/({d_{AB}^2+\sigma_A^2+\sigma_B^2}).$
Equivalently, $g$ is one divided by one plus the spread-to-separation ratio. It approaches one when separation dominates local spread, i.e., $d_{AB}^2\gg \sigma_A^2+\sigma_B^2$, equals $1/2$ when these squared scales are equal, and is small when the regions' spread dominates.

\begin{example}
Reconsider Fig.~\ref{fig:example-region}, where $C_1=\{p_1=(0.18,0.29),p_2=(1.11,0.13),p_3=(1.75,-0.11),p_6=(0.81,0.73)\}$, $C_2=\{p_5=(-0.17,0.91),p_9=(-0.28,1.84),p_{13}=(-0.09,2.81),p_{14}=(0.98,2.98)\}$ and $C_3=\{p_4=(3.04,0.28),p_7=(1.89,0.87),p_8=(2.89,0.70),p_{12}=(2.94,2.24)\}$. We have $c_{C_1}=(0.96,0.26)$, $c_{C_2}=(0.11,2.14)$, $\sigma_{C_1}^2=0.41$, and $\sigma_{C_2}^2=0.95$. Thus $ \lambda_{21}=(0.11,2.14)+\R(0.85,-1.88),$ $d_{21}^2=4.24$, and $g(C_2,C_1)=0.7573$. Similarly, we have $g(C_2,C_3)=0.8226$. Thus, $g(C_2,C_3)$ is larger than $g(C_2,C_1)$, which means the connection to $C_3$ spans more of the space relative to the two regions' extents.
\label{ex:region-score}
\end{example}

\stitle{Support for the Representative Secant:}
Any two distinct centroids define a line, even when their surrounding points are spread away from it. For each actual point $z\in A\cup B$, we therefore measure its squared perpendicular distance to $\lambda_{AB}$ relative to $d_{AB}^2$ and average its contribution: $ h(A,B)=\frac{1}{|A|+|B|}\sum_{z\in A\cup B}\frac{1}{1+\dist^2(z,\lambda_{AB})/d_{AB}^2}.$
A point on the secant contributes one; at perpendicular distance $d_{AB}$ it contributes $1/2$; and the contribution decreases smoothly with distance. Normalizing by $d_{AB}^2$ compares transverse deviation against the same longitudinal scale used by $g$. The average avoids preferring a pair merely because its leaves contain more points. Whereas $g$ uses total spread, $h$ specifically measures spread perpendicular to the proposed line.

\begin{figure}
  \subfigure[$C_2$ and $C_1$: $\lambda_{21}$]
  {
      \includegraphics[width=0.36\columnwidth]{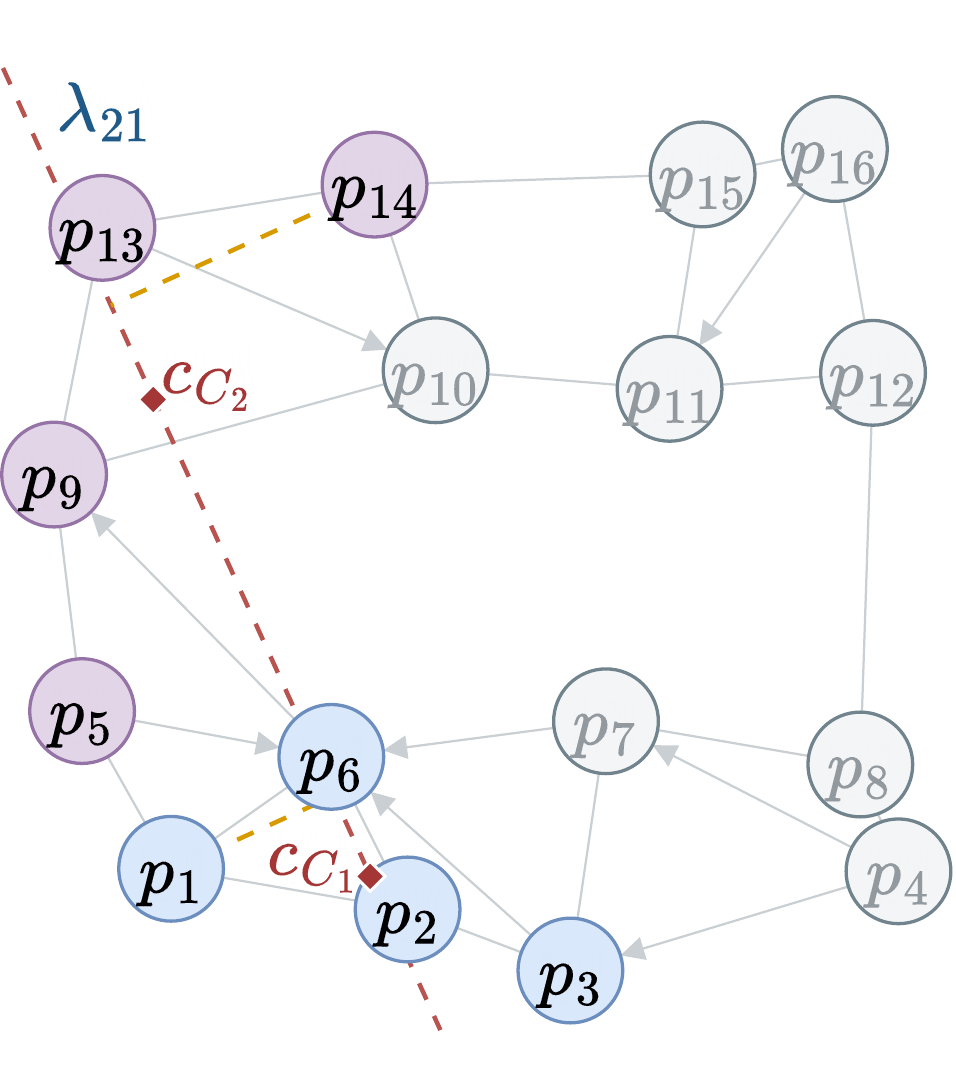}
      \label{fig:score-21}
  }
\hspace{4mm}
  \subfigure[$C_2$ and $C_3$: $\lambda_{23}$]
  {
      \includegraphics[width=0.4\columnwidth]{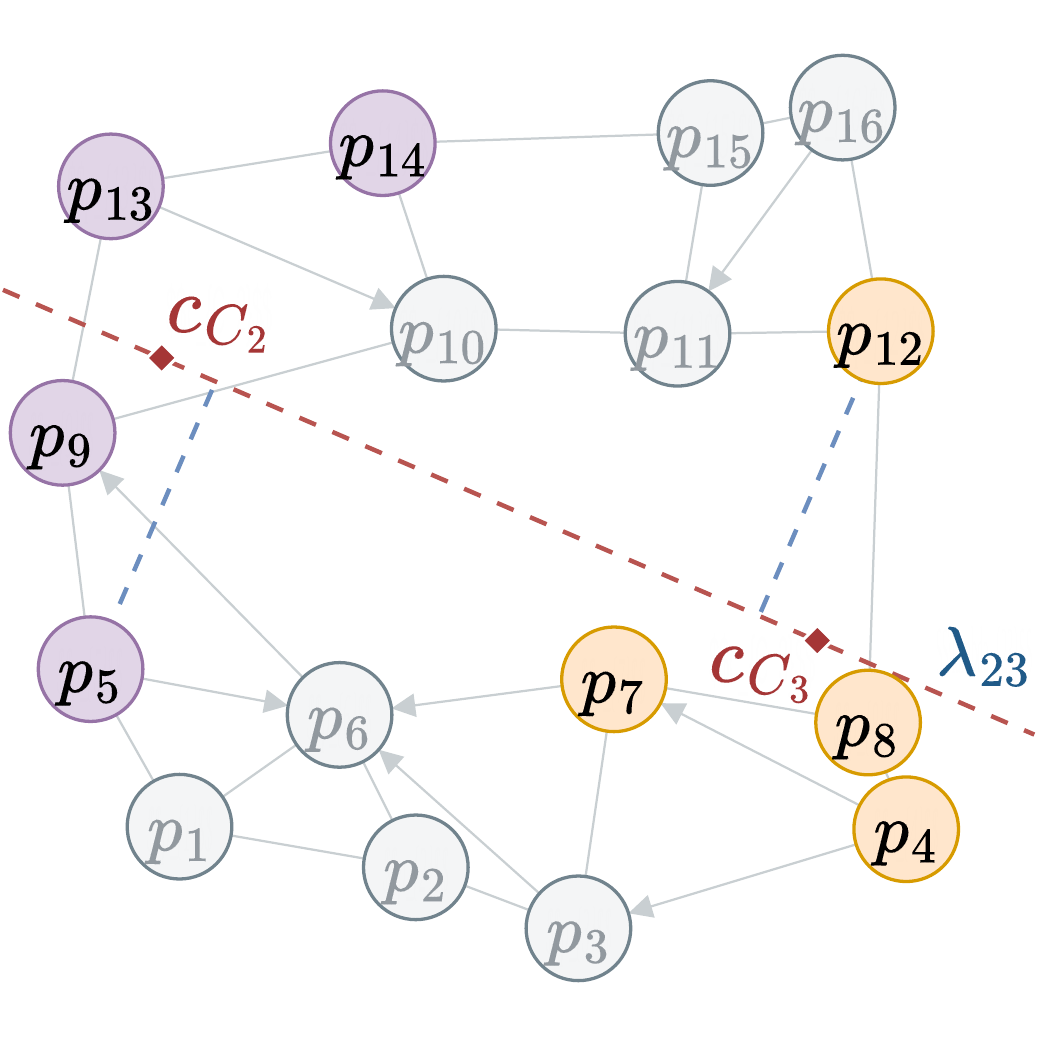}
        \label{fig:score-23}
  }

\vspace{-4mm}
\caption{An example of representative secants}
\label{fig:score-example}
\end{figure}

\begin{example}
\label{ex:score-residual}
Reconsider Example~\ref{ex:region-score}. Fig.~\ref{fig:score-example} shows the two centroids, their secant, and example perpendicular residuals.
In Fig.~\ref{fig:score-21}, $p_{13}$ and $p_{14}$ have perpendicular distances approximately $0.10$ and $1.14$ to this line. Thus, $p_{13}$ is much closer to $\lambda_{21}$ than $p_{14}$ and therefore contributes more, i.e., their contributions to $h$ are approximately $1.00$ and $0.77$. Summing the contributions of all eight points gives $7.42$; dividing by eight gives $h(C_2,C_1)=0.9270$. Similarly, in Fig.~\ref{fig:score-23}, we have $h(C_2,C_3)=0.9271$.
\label{ex:score-scent}
\end{example}

\stitle{Combining the Considerations:}
We retain the product $S(A,B)=g(A,B)\,h(A,B),$ setting $S(A,B)=0$ when the centroids coincide. Both terms lie in $[0,1]$, and $S$ cannot exceed either term. Thus, a large value requires both appreciable separation and secant support, rather than allowing one to compensate without limit for the other.

\begin{example}
\label{ex:score-combination}
Reconsider Examples~\ref{ex:region-score} and \ref{ex:score-scent} with $C_4=\{p_{10}=(1.22,2.25),p_{11}=(2.14,2.23),p_{15}=(2.27,3.02),p_{16}=(2.79,3.12)\}$.
We can compare all three possible target regions of $C_2$: for $C_1$, $g(C_2,C_1)=0.7573$, $h(C_2,C_1)=0.9270$, thus $S(C_2,C_1)=0.7020$;
for $C_3$, $g(C_2,C_3)=0.8226$ and $h(C_2,C_3)=0.9271$, thus $S(C_2,C_3)=0.7626$;
and for $C_4$, $g(C_2,C_4)=0.7470$ and $h(C_2,C_4)=0.9371$, thus $S(C_2,C_4)=0.7000$.
The scores show why support is not sufficient alone: $C_4$ gives the highest $h$ for source $C_2$, but its lower separation factor makes it a less preferred partner than $C_3$.
\end{example}

\stitle{Time Complexity:}
Each source has $O(s_hh+n_w)=O(n_w)$ candidates, since the hierarchy has $h=O(\log|\mathcal L|)$ levels. On a 64-bit machine, synchronous propagation takes $O\bigl(\kappa(|\mathcal L|+|E(Q)|)\lceil n_p/64\rceil\bigr)$ time~\cite{ThenKCHPK0V14}. Decoding proposals costs $O(|\mathcal L|n_p)$, each region score costs $O(bd)$, and retaining the best $r$ with a bounded heap costs $O(\log(r+1))$ per candidate. With cached leaf statistics and expected constant-time hash deduplication, Algorithm~\ref{algo:select-regions} thus takes $O\Bigl(\kappa(|\mathcal L|+|E(Q)|)\lceil n_p/64\rceil+|\mathcal L|n_w\bigl(bd+\log(r+1)\bigr)\Bigr)$.

\section{Search on the Line-Enhanced Graph}
\label{sec:search}

In this section, we introduce the search algorithm on the line-enhanced proximity graph.
We search $G^+$ based on the beam search introduced in Section~\ref{sec:pre}. The candidate pool has beam width $w\ge k$ and is ordered by point-to-line distance. During search, edges in $\Ereg$ are explored first; pair edges provide additional entrances when the current frontier does not already contain a closer continuation.

\begin{algorithm}[t]
\small
\SetVline
\SetFuncSty{textsf}
\SetArgSty{textsf}
\caption{\kw{SearchLPG}($G^+,L,k,w$)}
\label{algo:search}
\Input{line-enhanced graph $G^+$ with leaf-owner map $\ell$, query line $L$, result size $k$, and beam width $w\ge k$}
\Output{$k$ approximate nearest points to $L$}
$H\leftarrow\emptyset$; $\mathit{Visit}\leftarrow\emptyset$; $\mathit{Done}\leftarrow\emptyset$\;\nllabel{ln:search-init}
$p_0\leftarrow$ the entry point of $G^+$; $H\leftarrow\{p_0\}$; $\mathit{Visit}\leftarrow\{p_0\}$\;\nllabel{ln:search-entry}
\While{there is an unexpanded point in $H$\nllabel{ln:search-loop}}{
 $p\leftarrow$ the closest unexpanded point to $L$ in $H$\;\nllabel{ln:search-pop}
 Mark $p$ expanded; $\mathit{Done}\leftarrow\mathit{Done}\cup\{\ell(p)\}$\;\nllabel{ln:search-done}
 \For{each $(p,y)\in\Ereg$\nllabel{ln:search-ordinary}}{
  \lIf{$y\notin\mathit{Visit}$}{$H\leftarrow H\cup \{y\}$; $\mathit{Visit}\leftarrow \mathit{Visit}\cup \{y\}$;}\nllabel{ln:search-ordinary-eval}
 }
 \textbf{while} $|H|>w$ \textbf{do} remove the farthest point from $L$ in $H$\;\nllabel{ln:search-prune}
  $F\leftarrow$ the unexpanded points in $H$\;\nllabel{ln:search-frontier}
 \lIf{some $z\in F$ satisfies $\dist(z,L)<\dist(p,L)$}{\textbf{continue}}\;\nllabel{ln:search-progress}
 \For{each $(p,y)\in\Epair$ in stored order\nllabel{ln:search-pairs}}{
  \If{$y\notin\mathit{Visit}$ and $\ell(y)\notin\mathit{Done}$\nllabel{ln:search-pair-eval}}{$H\leftarrow H\cup \{y\}$; $\mathit{Visit}\leftarrow \mathit{Visit}\cup \{y\}$;\nllabel{ln:search-pair-add}}
 }
 \textbf{while} $|H|>w$ \textbf{do} remove the farthest point from $L$ in $H$\;\nllabel{ln:search-prune-2}
}
\Return{the $k$ closest points to $L$ in $H$\nllabel{ln:search-return}}
\end{algorithm}

\stitle{Local Exploration:}
Let $H$ be the candidate pool, $\mathit{Visit}$ the points whose distances have been evaluated, and $\mathit{Done}$ the leaves containing an expanded point (line~\ref{ln:search-init}).
The pool $H$ keeps at most $w$ points, including expanded points; each newly retained point is unexpanded.
Initially, we evaluate an entry point $p_0$ and insert it into $H$ and $\mathit{Visit}$ (line~\ref{ln:search-entry}).
Each iteration selects the closest unexpanded point $p\in H$ (line~\ref{ln:search-pop}), marks it expanded (line~\ref{ln:search-done}), and examines its retained original neighbors (lines~\ref{ln:search-ordinary}-\ref{ln:search-ordinary-eval}).
The farthest point from $L$ will be iteratively removed until the size of $H$ is no more than $w$ (line~\ref{ln:search-prune}).
Let $F$ be the unexpanded points now in $H$ (line~\ref{ln:search-frontier}). If some $z\in F$ satisfies $\dist(z,L)<\dist(p,L)$, the frontier already offers progress and we skip the pair edges of $p$ for this expansion (line \ref{ln:search-progress}).

\stitle{Exploration through Pair Edges:}
Otherwise, we consider the pair endpoints $x$ of $p$ in stored order, i.e., ordered by decreasing $J(p,x;\ell(x))$ as discussed in Section~\ref{sec:overview} (line~\ref{ln:search-pairs}).
An endpoint is skipped if its distance is already known or its owner leaf belongs to $\mathit{Done}$. The leaf test avoids repeated entrances into a region already being explored. Other endpoints are evaluated and inserted into $H$ (line \ref{ln:search-pair-add}).
Similarly, we iteratively remove the farthest nodes from $L$ in $H$ to ensure $|H|\le w$ (line~\ref{ln:search-prune-2}).
Finally, the $k$ closest points to $L$ in $H$ are returned (line~\ref{ln:search-return}).

\begin{example}
\label{ex:search}
Reconsider Fig.~\ref{fig:motivation-linePG} with $p_0=p_9$ and $w=k=3$. After considering the retained original neighbors $p_5,p_{13}$, the closest unexpanded candidate is at distance approximately $0.70$, larger than $\dist(p_9,L)\approx0.12$. Pair exploration is thus allowed. Since $C_3$ has not been expanded, $p_8$ enters the pool at distance approximately $0.24$. Its original neighbors subsequently provide $p_4,p_7$, yielding $\{p_7,p_4,p_9\}$.
\end{example}

\section{Experiments}
\label{sec:exp}

\noindent\textbf{Datasets:}
\label{sec:settings}
We employ 6 real-world public datasets for evaluation. The datasets are summarized in Table~\ref{tab:dataset}. Each query contains two $d$-dimensional endpoints encoding one infinite line.

\yaleb and \multipie provide controlled variation in face appearance. \yaleb release documents 28 subjects, nine poses, and 64 illumination conditions~\cite{GeorghiadesBK01,LeeHK05}; Multi-PIE documents 337 subjects, multiple sessions, 15 viewpoints, and 19 illumination conditions~\cite{GrossMCKB10}. 
For both \yaleb and \multipie, we generate their vector representations using a 128-dimensional principal component analysis embedding, following the classical eigenface representation~\cite{jocn.1991.3.1.71}.
For \yaleb, its queries are generated by considering varying illumination while holding the subject and pose fixed. That is, for each subject--pose group, we select two images as observations under different illumination conditions, define a line through their embeddings as a query. Using one pair per group and excluding all 504 endpoint images leaves 15{,}624 base vectors and yields 252 query lines.
For \multipie, its queries are generated by varying illumination while holding the subject, recording session, viewpoint, and expression fixed. Specifically, for each subject--session--viewpoint--expression group, we select two images acquired under different illumination conditions as the query endpoints. Using one pair per group and excluding all 12{,}948 distinct endpoint images leaves 246{,}012 base vectors and yields 6{,}474 query lines.

\begin{table}[t]
\centering\small
\caption{Statistics of Datasets}
\vspace*{-3mm}
\label{tab:dataset}
\begin{tabular}{|l|r|r|r|c|}\hline \textbf{Datasets} & \textbf{$N$} & \textbf{$d$} & \textbf{\#queries} & \textbf{Source}\\ \hline \hline
\yaleb & 15,624 & 128 & 252 & \cite{GeorghiadesBK01,LeeHK05}\\ \hline
\celebA & 60,000 & 512 & 1,000 &\cite{KarrasALL18}\\ \hline
\ffhq & 70,000 & 512 & 1,000 & \cite{KarrasLA19}\\ \hline
\multipie & 246,012 & 128 & 6,474 & \cite{GrossMCKB10}\\ \hline
\SIFT & 1,000,000 & 128 & 10,000 & \cite{sift}\\ \hline
\GIST & 1,000,000 & 960 & 10,000 & \cite{sift}\\ \hline
\end{tabular}
\vspace*{-3mm}
\end{table}

\begin{figure*}
\includegraphics[width=0.8\columnwidth]{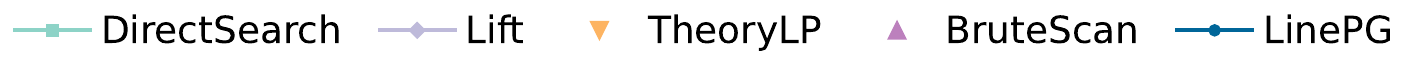}

\subfigure[{\yaleb} ($k=10$)]
{
    \includegraphics[width=0.31\columnwidth]{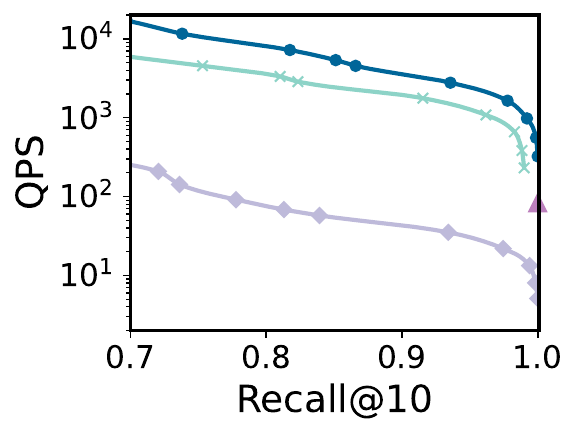}
}
\hfill
\subfigure[{\celebA} ($k=10$)]
{
    \includegraphics[width=0.31\columnwidth]{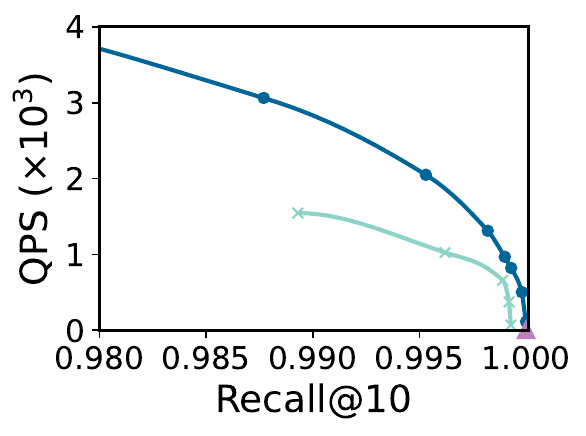}
}
\hfill
\subfigure[{\ffhq} ($k=10$)]
{
    \includegraphics[width=0.31\columnwidth]{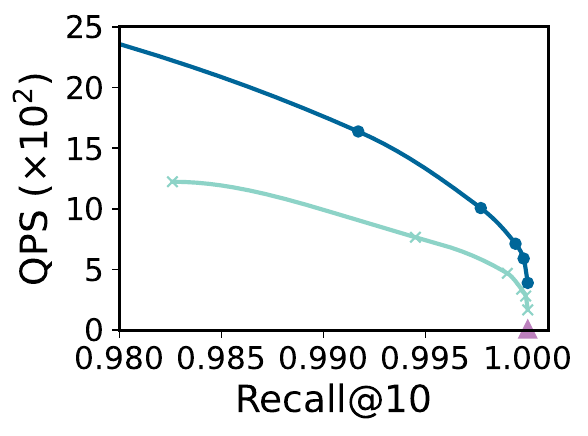}
}
\hfill
\subfigure[{\multipie} ($k=10$)]
{
    \includegraphics[width=0.31\columnwidth]{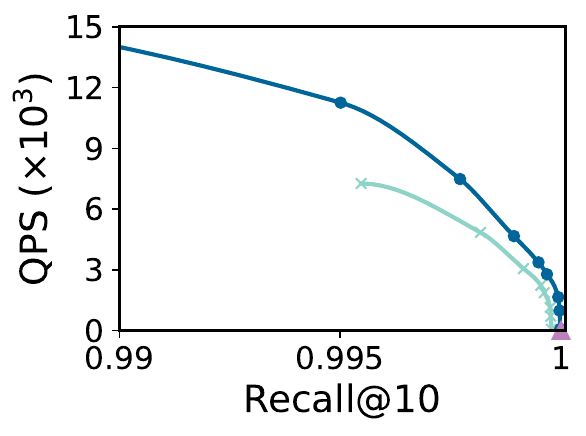}
}
\hfill
\subfigure[{\SIFT} ($k=10$)]
{
    \includegraphics[width=0.31\columnwidth]{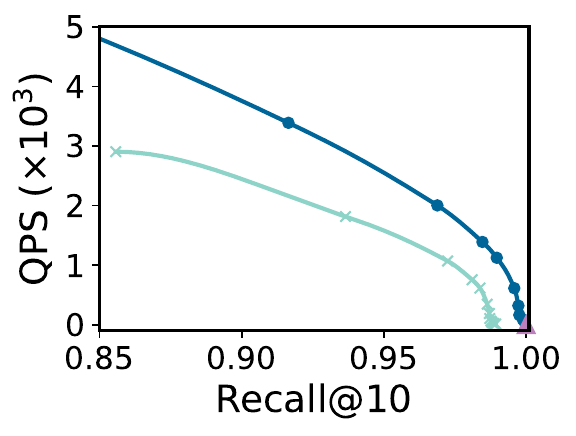}
}
\hfill
\subfigure[{\GIST} ($k=10$)]
{
    \includegraphics[width=0.31\columnwidth]{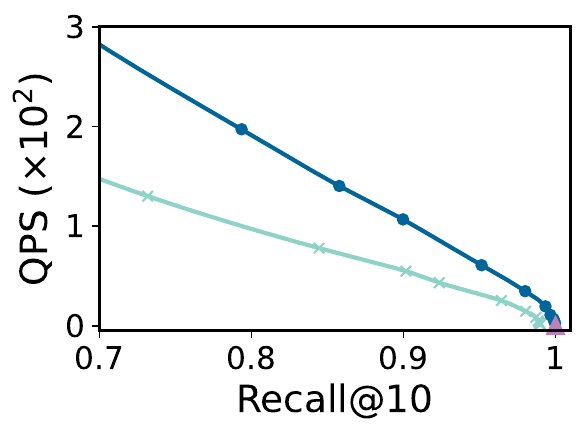}
}

\subfigure[{\yaleb} ($k=100$)]
{
    \includegraphics[width=0.31\columnwidth]{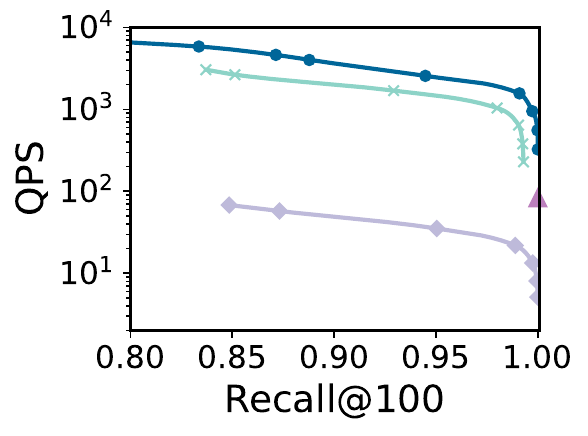}
}
\hfill
\subfigure[{\celebA} ($k=100$)]
{
    \includegraphics[width=0.31\columnwidth]{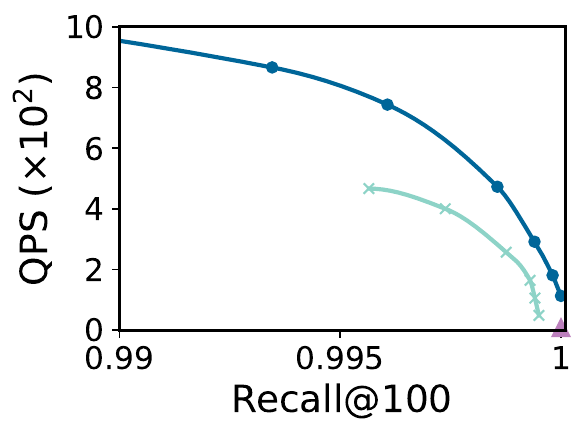}
}
\hfill
\subfigure[{\ffhq} ($k=100$)]
{
    \includegraphics[width=0.31\columnwidth]{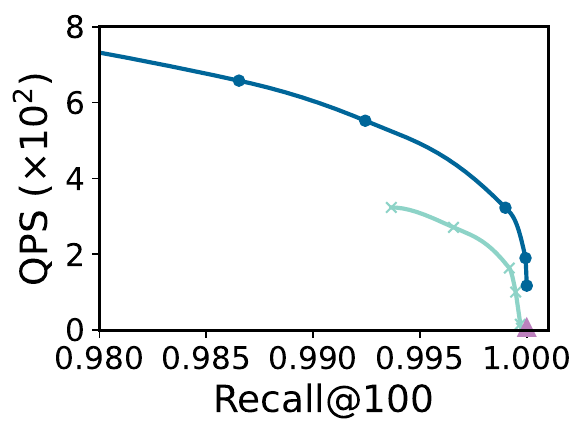}
}
\hfill
\subfigure[{\multipie} ($k=100$)]
{
    \includegraphics[width=0.31\columnwidth]{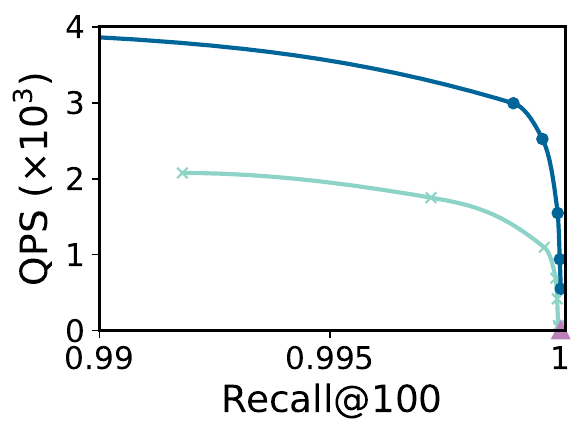}
}
\hfill
\subfigure[{\SIFT} ($k=100$)]
{
    \includegraphics[width=0.31\columnwidth]{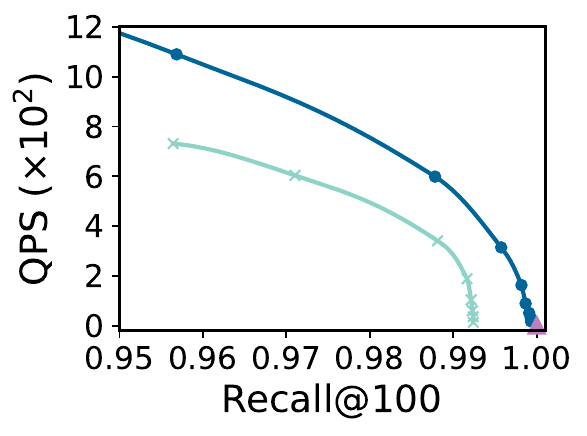}
}
\hfill
\subfigure[{\GIST} ($k=100$)]
{
    \includegraphics[width=0.31\columnwidth]{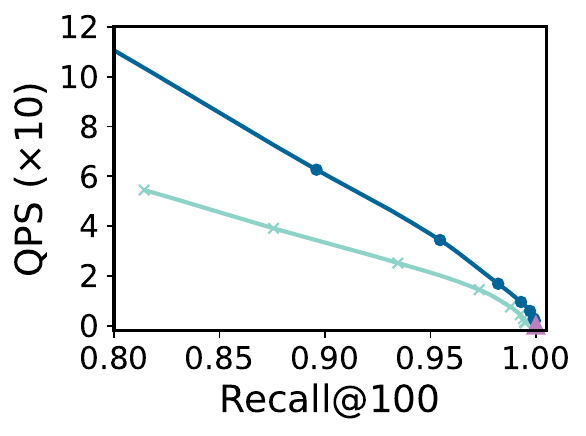}
}
\vspace{-3mm}
\caption{Search performance comparisons on full datasets (Exp. 1)}

\vspace{-3mm}
\label{fig:search-overall}
\end{figure*}

\begin{figure*}
\includegraphics[width=0.8\columnwidth]{figures/qps-512-legend}

\subfigure[{\yaleb} ($k=10$)]
{
    \includegraphics[width=0.31\columnwidth]{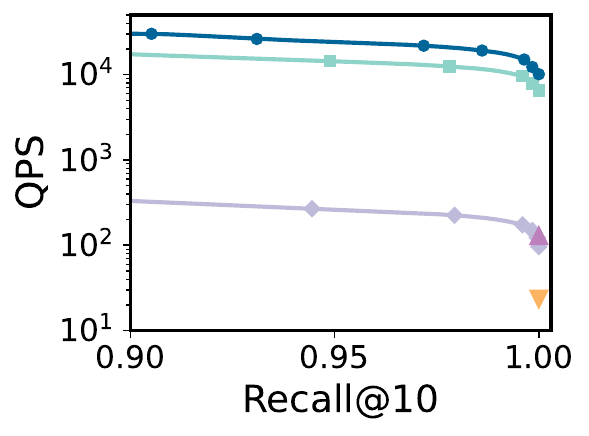}
}
\hfill
\subfigure[{\celebA} ($k=10$)]
{
    \includegraphics[width=0.31\columnwidth]{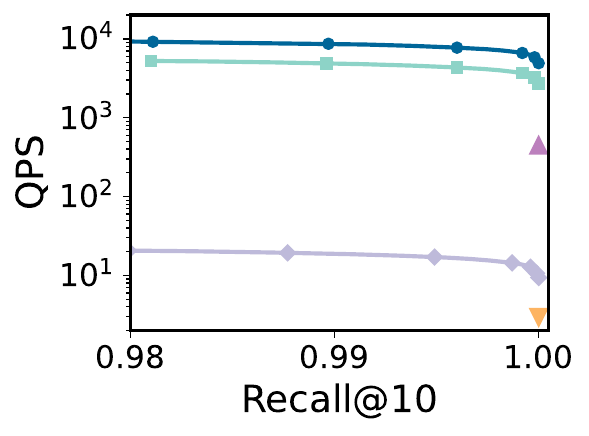}
}
\hfill
\subfigure[{\ffhq} ($k=10$)]
{
    \includegraphics[width=0.31\columnwidth]{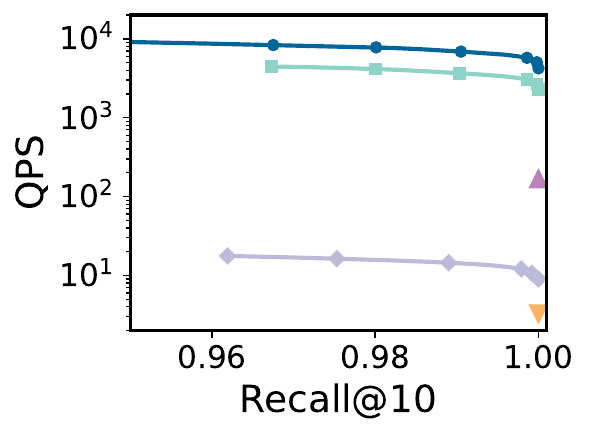}
}
\hfill
\subfigure[{\multipie} ($k=10$)]
{
    \includegraphics[width=0.31\columnwidth]{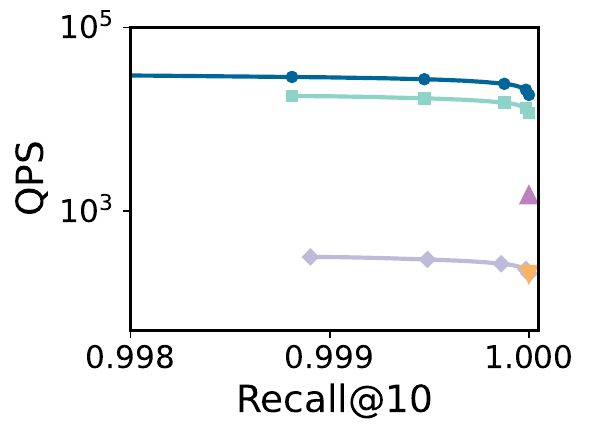}
}
\hfill
\subfigure[{\SIFT} ($k=10$)]
{
    \includegraphics[width=0.31\columnwidth]{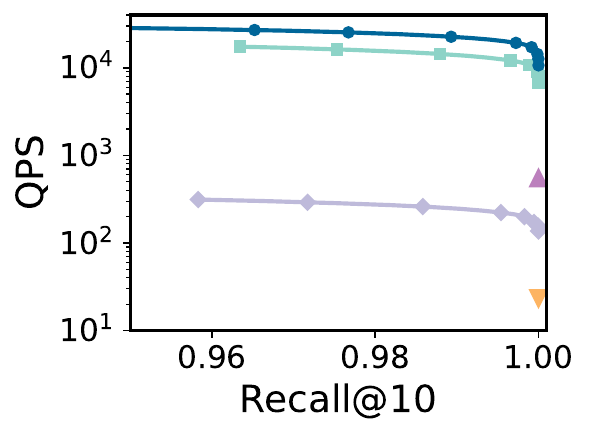}
}
\hfill
\subfigure[{\GIST} ($k=10$)]
{
    \includegraphics[width=0.31\columnwidth]{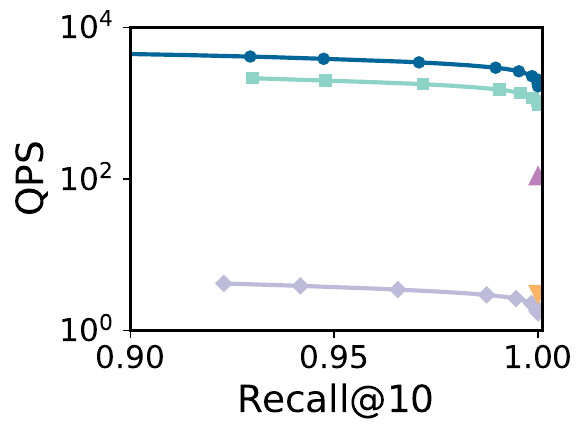}
}
\vspace{-3mm}
\caption{Search performance comparisons on $N=512$ subsets (Exp. 1)}

\vspace{-4mm}
\label{fig:search-512}
\end{figure*}

\celebA is introduced with progressive GAN training~\cite{KarrasALL18}, and \ffhq is introduced with StyleGAN~\cite{KarrasLA19}. 
For both datasets, we use the e4e inversion encoder~\cite{TovANPC21} to map real images into the latent space of a pretrained StyleGAN model and derive a 512-dimensional representation from each inverted code.
For these two datasets, their queries are constructed from an image representation $z$ and a nonzero semantic direction $u$ in the same representation space, motivated by approximately linear facial-attribute editing in GAN latent spaces~\cite{ShenGTZ20}.
Specifically, the endpoints $z$ and $z+u$ encode the query line $\{z+\lambda u\mid\lambda\in\mathbb{R}\}$.
We generate 1{,}000 query lines for each, providing additional real-image workloads for evaluating retrieval along semantic directions in a learned latent space.

\SIFT and \GIST are standard vector-search benchmarks consisting of image descriptors and are widely used in point ANN search studies~\cite{WangXY021,taumg}.
We retain the released base vectors and construct line queries from the released point-query sets.
Specifically, we uniformly sample 10{,}000 pairs of query vectors and define each query line as the infinite line passing through the sampled pair.

\noindent\textbf{Algorithms}:
We compare our approach, named \ours, with four baselines: \Theory~\cite{AndoniIKN09}, \Lift, \DirSearch, and \Brute. For each algorithm, we try to tune the parameters for index construction for a Pareto-optimal recall-QPS curve. For fair comparisons, the graph-based approaches, i.e., \Lift, \DirSearch and \ours, use HNSW~\cite{hnsw} with the same construction settings, $M=32$ and ${efC}=500$, and match the insertion order and construction seed.

\textbf{(1)} \Theory implements the algorithm introduced in ~\cite{AndoniIKN09}, with dense-group size $m=\lceil\sqrt N\rceil$, approximation parameter $\epsilon=0.5$, exponent $t=1/8$, and projection dimension $\min\{d,\lceil2/t\rceil+1\}=\min\{d,17\}$. To support top-$k$ retrieval, we execute one native query pass and retain the $k$ closest points, under the original point-to-line distance, among all evaluated database points. We include \Theory only in the small-base comparison because of its construction cost.
\textbf{(2)} \DirSearch uses the original proximity graph $G$ and evaluates candidates via beam search with point-to-line distance, as introduced in Section~\ref{sec:limit}.
\textbf{(3)} \Lift uses the vector representation with $D=d(d+1)/2+d$ coordinates for index construction and search, as introduced in Section~\ref{sec:limit}. It builds a separate HNSW index over the lifted vectors, while using the same HNSW construction parameters as \DirSearch.
\textbf{(4)} \Brute directly evaluates $\dist^2(p,L)$ for every $p\in\Pset$ and returns the $k$ smallest distances. It has no index or search-width parameter and provides an exact-search timing reference at recall one.
\textbf{(5)} \ours enhances the same $G$ used by \DirSearch, with total degree limit $M=32$ and pair-edge limit $\Mpair=4$. Unless stated otherwise, we set the leaf capacity $b=32$, smoothing rounds $t_s=4$, sampling budget $s_h=4$, quotient-hop radius $\kappa=3$, and retained region partners $r=16$. Each representative bank of leaf nodes contains up to eight members nearest its centroid and up to four farthest remaining members.

\noindent\textbf{Performance metrics:}
Following existing vector search works~\cite{WangXY021,abs-2410-01231,taumg,LiYL26,HuangLT21}, we employ QPS (Queries Per Second) to measure efficiency and use Recall to show the accuracy. QPS is the quantity of queries processed per second, while Recall is defined in Section~\ref{sec:pre}.

\noindent\textbf{Experimental environments}:
The experiments are conducted on a CentOS Linux server with Intel(R) Xeon(R) Silver 4215 CPU @ 2.50GHz with 128GB memory. All algorithms are implemented in C++17. The code is compiled with g++ 11.5.0 under O3 optimization.

For index construction, we report the index time and size on the same machine using multiple threads. For vector search, we set $k=10$ by default and evaluate the performance of the search algorithms using a single thread following~\cite{ALMG,taumg,HuangLT21,WangXY021}, and incrementally adjust parameters within the search algorithm to achieve the desired recall in the experiments.

\stitle{Exp. 1: Search Performance Comparison.}
Figs.~\ref{fig:search-overall} and~\ref{fig:search-512} compare \ours with \DirSearch, \Lift, and \Brute on the full datasets with $k=\{10,100\}$ and the $N=512$ subsets with $k=10$, respectively. The results show that \ours improves query efficiency at high recall while also extending the recall attainable by direct graph search.
\Theory is evaluated only on the small bases due to its high index construction cost, as shown in Exp. 2.
For full-set \Lift, construction times out after two hours on \celebA, \SIFT and \GIST, and runs out of memory on \ffhq. It completes on \yaleb and \multipie, as reported in Table~\ref{tab:index-comparison}.

\begin{table*}[t]
  \centering \small
  \caption{Index construction cost comparisons (Exp. 2)}
  \vspace*{-3mm}
  \label{tab:index-comparison}
  \resizebox{0.92\linewidth}{!}{
  \begin{tabular}{|c|l|r|r|r|r|r|r|r|r|r|r|r|}
  \hline
  \multirow{3}{*}{\textbf{Scale}}
  & \multirow{3}{*}{\textbf{Dataset}}
  & \multicolumn{7}{c|}{\textbf{Index Construction Time (s for Full, ms for $N=512$)}}
  & \multicolumn{4}{c|}{\textbf{Index Storage (MiB for Full, KiB for $N=512$)}}\\
  \cline{3-13}
  & & \multirow{2}{*}{\Theory}
  & \multirow{2}{*}{\DirSearch}
  & \multirow{2}{*}{\Lift}
  & \multicolumn{4}{c|}{\ours}
  & \multirow{2}{*}{\Theory}
  & \multirow{2}{*}{\DirSearch}
  & \multirow{2}{*}{\Lift}
  & \multirow{2}{*}{\ours}\\
  \cline{6-9}
  & & & & & \textbf{Total}
  & {H+Q} & {Region}
  & {Edge} & & & & \\
  \hline\hline

  \multirow{6}{*}{\rotatebox[origin=c]{90}{Full}}
  & \yaleb
  & -- & \deepcolor 0.630 & 177.758 & \lightcolor 4.957
  & 0.168 & 0.070 & 4.089
  & -- & \deepcolor 8.799 & 500.86 & \lightcolor 9.199 \\ \cline{2-13}

  & \multipie
  & -- & \deepcolor 21.829 & 5,827.460 & \lightcolor 87.821
  & 5.685 & 1.061 & 59.246
  & -- & \deepcolor 143.334 & 7,890.53 & \lightcolor 148.503 \\ \cline{2-13}

  & \ffhq
  & -- & \deepcolor 44.592 & \textsc{oom} & \lightcolor 74.171
  & 2.100 & 0.458 & 27.021
  & -- & \deepcolor 147.348 & \textsc{oom} & \lightcolor 147.408 \\ \cline{2-13}

  & \celebA
  & -- & \deepcolor 17.967 & \textsc{oot} & \lightcolor 38.907
  & 1.344 & 0.255 & 19.341
  & -- & \deepcolor 124.468 & \textsc{oot} & \lightcolor 125.138 \\ \cline{2-13}

  & \SIFT
  & -- & \deepcolor 366.553 & \textsc{oot} & \lightcolor 721.441
  & 80.920 & 11.645 & 262.323
  & -- & \deepcolor 642.201 & \textsc{oot} & \lightcolor 642.241 \\ \cline{2-13}

  & \GIST
  & -- & \deepcolor 2,271.748 & \textsc{oot} & \lightcolor 2,527.428
  & 51.249 & 13.015 & 191.416
  & -- & \deepcolor 3,767.079 & \textsc{oot} & \lightcolor 3,773.679 \\
  \hline

  \multirow{6}{*}{\rotatebox[origin=c]{90}{$N=512$}}
  & \yaleb
  & 144,413 & \deepcolor 19.93 & 8,350 & \lightcolor 61.89
  & 4.35 & 4.02 & 33.58
  & 465.51 & \deepcolor 290.91 & 17,075.93 & \lightcolor 303.28 \\ \cline{2-13}

  & \multipie
  & 144,242 & \deepcolor 19.21 & 8,262 & \lightcolor 50.91
  & 4.34 & 3.19 & 24.18
  & 465.59 & \deepcolor 293.88 & 17,079.24 & \lightcolor 305.05 \\ \cline{2-13}

  & \ffhq
  & 559,685 & \deepcolor 34.11 & 146,241 & \lightcolor 61.63
  & 12.29 & 4.03 & 11.20
  & 1,416.59 & \deepcolor 1,066.62 & 264,718.30 & \lightcolor 1,071.17 \\ \cline{2-13}

  & \celebA
  & 559,078 & \deepcolor 34.90 & 144,678 & \lightcolor  74.91
  & 12.29 & 3.91 & 23.81
  & 1,416.56 & \deepcolor 1,063.35 & 264,715.50 & \lightcolor 1,073.64 \\ \cline{2-13}

  & \SIFT
  & 144,485 & \deepcolor 19.64 & 8,992 & \lightcolor 51.02
  & 5.09 & 3.73 & 22.56
  & 465.59 & \deepcolor 295.43 & 17,079.29 & \lightcolor 306.29 \\ \cline{2-13}

  & \GIST
  & 983,791 & \deepcolor 52.25 & 447,206 & \lightcolor 90.40
  & 19.90 & 3.43 & 14.82
  & 2,526.09 & \deepcolor 1,961.37 & 926,430.33 & \lightcolor 1,965.77 \\
  \hline
  \end{tabular} }

\end{table*}

On the full datasets, we first compare QPS when recall is at least $0.99$. For $k=10$, \ours achieves QPS speedups of $4.23$x, $2.00$x, $2.14$x, and $1.55$x over \DirSearch on \yaleb, \celebA, \ffhq, and \multipie, respectively. For $k=100$, the speedups are $2.44$x, $1.86$x, $1.71$x, $1.44$x, $1.67$x, and $2.15$x on \yaleb, \celebA, \ffhq, \multipie, \SIFT, and \GIST, respectively. \DirSearch does not reach $0.99$ recall on full \SIFT or \GIST for $k=10$ within the tested search range.
Importantly, the efficiency advantage is not obtained by sacrificing retrieval quality. For example, \ours attains reported recall $1.0$ for both values of $k$ on full \yaleb, \celebA, \ffhq, and \multipie.
On \SIFT, its highest observed recalls are $0.9993$ and
$0.9994$ for $k=10$ and $100$, respectively, whereas
\DirSearch reaches only $0.9893$ and $0.9924$.
On \GIST, the highest observed recalls of \ours are $0.9997$ and
$0.9995$ for $k=10$ and $100$, respectively, whereas \DirSearch
reaches only $0.9899$ and $0.9948$.
Moreover, higher efficiency and higher recall can be achieved simultaneously, e.g., on \SIFT with $k=10$, \ours obtains $0.9958$ recall at $613.11$ QPS, while the highest-recall tested configuration of \DirSearch obtains $0.9893$ at $13.70$ QPS. This is $44.76$x higher throughput with better recall, not an equal-recall interpolation.
At the same reported recall of $1.0$ on full \yaleb, \ours also outperforms \Lift by $63.18$x and $63.34$x for $k=10$ and $100$, respectively.

On the $N=512$ subsets, \ours reaches reported recall $1.0$ on all six datasets.
At this same recall, its QPS is $1.56$x, $1.81$x, $1.85$x, $1.57$x, $1.57$x, and $1.76$x that of \DirSearch on \yaleb, \celebA, \ffhq, \multipie, \SIFT, and \GIST, respectively.
Its speedups over \Lift range from $77.73$x on \SIFT to $976.17$x on \GIST, with the same recall.
Compared with \Theory, \ours\ is $431.80$x, $1{,}665.63$x, $604.74$x, $90.15$x, $257.92$x, and $551.81$x faster on \yaleb, \celebA, \ffhq, \multipie, \SIFT and \GIST, respectively, again at reported recall $1.0$.
Together, these results demonstrate a favorable high-recall QPS tradeoff on both the full and small bases.

\stitle{Exp. 2: Index Size and Construction Time.}
Table~\ref{tab:index-comparison} compares the index sizes and construction times of different approaches in Exp. 1, where ``--'' in the table denotes \Theory is not tested, \textsc{oom} and \textsc{oot} denote the out-of-memory failures and timeouts as discussed. 

For index time, Table~\ref{tab:index-comparison} further breaks \ours's construction into HNSW (the time is the same as \DirSearch), hierarchy and quotient construction (denoted as H+Q), candidate region pair generation (denoted as Region), and pair edge selection and addition (denoted as Edge). Compared with \DirSearch, on \GIST, these latter three stages add $255.680$\,s to the $2{,}271.748$\,s HNSW build, for a total of $2{,}527.428$\,s. Pair edge selection dominates these three additional stages, accounting for $73.9$--$94.5\%$ across the full datasets. Compared with two other baselines, the index construction time of \ours is significantly smaller than \Lift and \Theory.

For index size, we report the total storage required for retrieval, including the original base vectors and the method-specific
index structures. Since \DirSearch uses the original HNSW graph, we report its HNSW size. Our method enhances the HNSW; Table~\ref{tab:index-comparison} reports its total index size, including HNSW, changed edge lists, and leaf-owner labels.
On the full datasets, \ours occupies $1.000$--$1.046$ times the total storage of \DirSearch, e.g., its \SIFT totals are slightly larger with an almost zero increase. At $N=512$, the indices of \ours are much smaller than \Theory and \Lift.

\begin{figure}
\includegraphics[width=0.85\columnwidth]{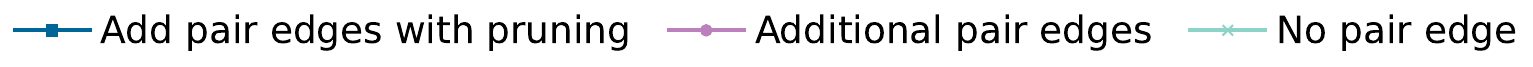}

\subfigure[{\yaleb}]
{
    \includegraphics[width=0.3\columnwidth]{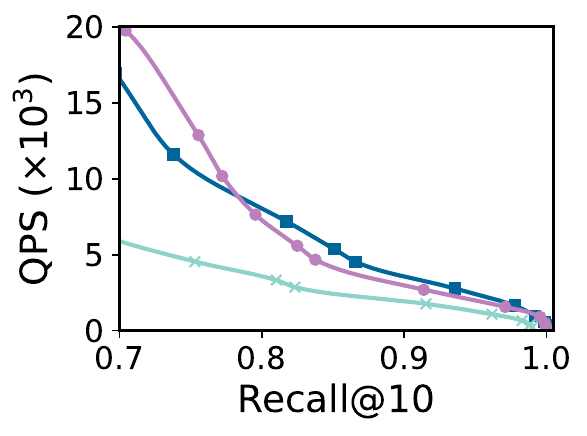}
}
\hfill
\subfigure[{\multipie}]
{
    \includegraphics[width=0.3\columnwidth]{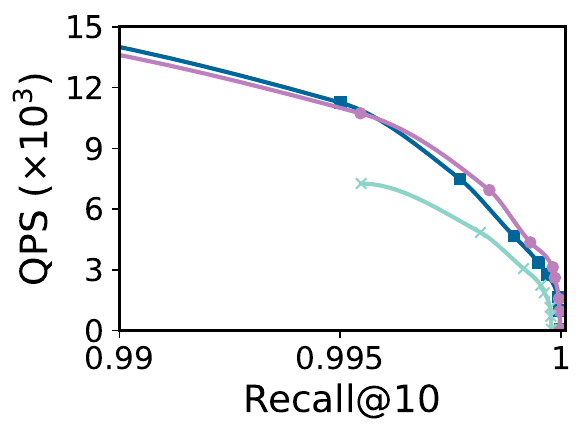}
}
\hfill
\subfigure[{\SIFT}]
{
    \includegraphics[width=0.3\columnwidth]{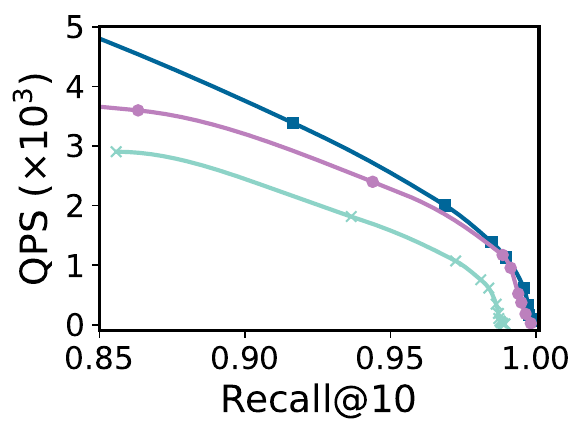}
}
\vspace{-3mm}
\caption{Ablation study for pair edges in indices (Exp. 3)}

\vspace{-4mm}
\label{fig:ablation-edge}
\end{figure}

\stitle{Exp. 3: Ablation Studies.}
We evaluate the following three ablations on the full \yaleb, \multipie, and \SIFT datasets with $k=10$.

\noindent\underline{Exp. 3-1: Effectiveness of Pair Edges.}
Fig.~\ref{fig:ablation-edge} compares three approaches:
\ding{192} Add pair edges with pruning (\ours), which inserts at most $\Mpair=4$ pair edges per point and prunes original edges when the total degree would exceed $M$;
\ding{193} Additional pair edges, which uses the same pair-edge budget without removing original edges, allowing a degree up to $M+\Mpair$; and
\ding{194} No pair edge (\DirSearch), which retains the original graph.

On \SIFT, \ding{194} reaches a maximum recorded recall of $0.9893$, compared with $0.9982$ for \ding{193} and $0.9993$ for \ding{192}. The two enhanced variants thus extend the measured high-recall range.
The benefit of retaining every original edge depends on the operating point.
At the $0.99$ recall target, \ding{192} is faster than \ding{193} on \yaleb and \multipie, whereas \ding{193} is faster on \SIFT.
This advantage reverses at the $0.995$ target on \SIFT, where \ding{192} achieves $3.38$x the QPS of \ding{193}.
Pruning is also faster at the same recorded recall of $1.0$ on \yaleb.
Thus, neither strategy consistently dominates, but both improve upon the original graph, and replacement maintains the degree bound without preventing high-recall retrieval.
For \ding{193}, construction takes $4.961$, $87.816$, and $726.490$ seconds, with index sizes of $9.57$, $155.29$, and $647.76$ MiB on \yaleb, \multipie, and \SIFT, respectively.
The corresponding costs of \ding{192} and \ding{194} are reported in Table~\ref{tab:index-comparison}.

\begin{figure}
\includegraphics[width=0.9\columnwidth]{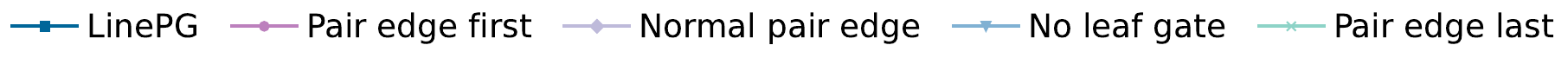}

\subfigure[{\yaleb}]
{
    \includegraphics[width=0.3\columnwidth]{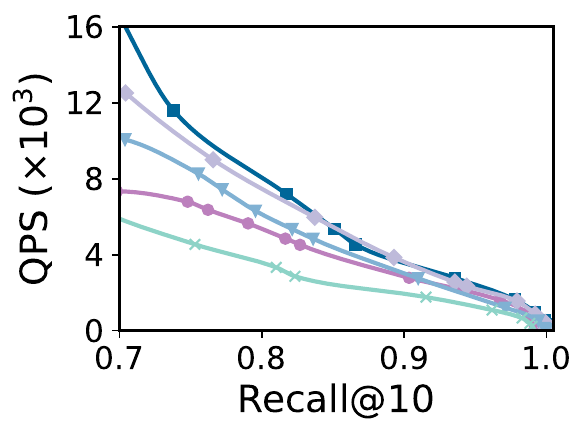}
}
\hfill
\subfigure[{\multipie}]
{
    \includegraphics[width=0.3\columnwidth]{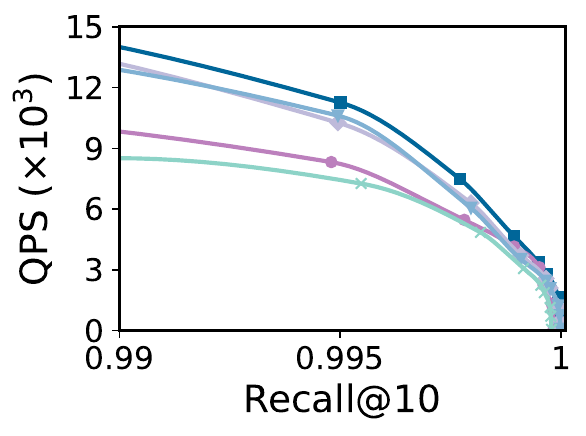}
}
\hfill
\subfigure[{\SIFT}]
{
    \includegraphics[width=0.3\columnwidth]{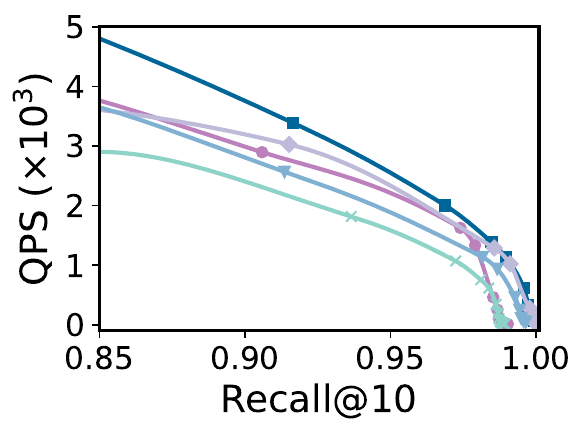}
}
\vspace{-3mm}
\caption{Ablation study for search algorithm (Exp. 3)}

\vspace{-4mm}
\label{fig:ablation-search}
\end{figure}

\noindent\underline{Exp. 3-2: Online Search Strategy.}
Fig.~\ref{fig:ablation-search} compares \ours with \ding{192} pair edge first, which explores edges in $\Epair$ before edges in $\Ereg$; \ding{193} normal pair edge, which treats pair edges as ordinary
adjacency entries; \ding{194} pair edge last, which explores edges in $\Ereg$ before edges in $\Epair$; and \ding{195} no leaf gate, which removes the test $\ell(y)\notin\mathit{Done}$ while retaining point-level visited checks.
These variants examine how pair edges are scheduled and whether
repeated entrances into an already explored leaf should be suppressed.

Compared with \ding{192} pair edge first and \ding{194} pair edge last, \ours always has better search performance, which supports selective interleaving of local exploration and inter-region probes.
The comparison with \ding{193} normal pair edge nevertheless reveals a tradeoff, rather than uniform dominance.
On \SIFT, \ours achieves $2.23$x its QPS at a recall
target of $0.995$. At a target of $0.999$, however, \ding{193} attains
recall $0.9995$ at $128.16$ QPS, compared with recall $0.9993$ at $61.31$ QPS for \ours, making \ding{193} $2.09$x faster.
Thus, selective pair-edge use improves efficiency at several
measured operating points but unrestricted exploration remains competitive in the high-recall tail.
Compared with \ding{195} no leaf gate, the check to avoid repeat region exploration leads to a higher QPS at similar recall.

\begin{figure}
\includegraphics[width=0.8\columnwidth]{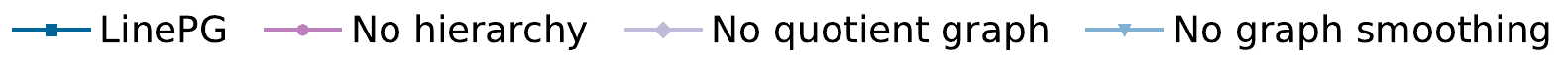}

\subfigure[{\yaleb}]
{
    \includegraphics[width=0.3\columnwidth]{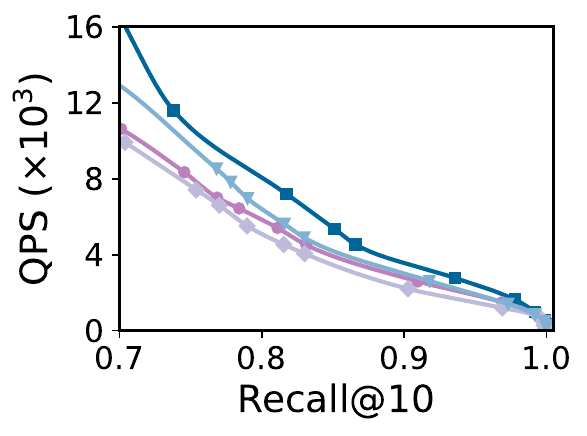}
}
\hfill
\subfigure[{\multipie}]
{
    \includegraphics[width=0.3\columnwidth]{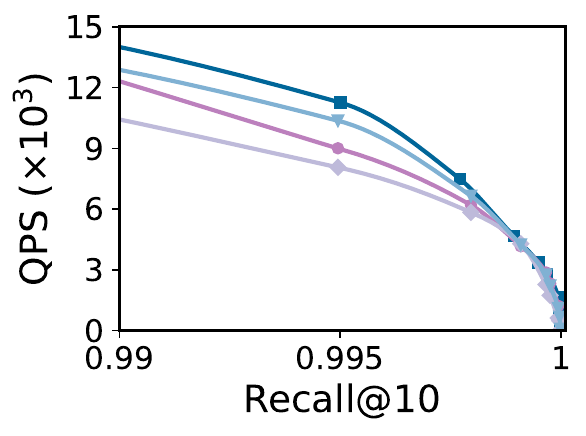}
}
\hfill
\subfigure[{\SIFT}]
{
    \includegraphics[width=0.3\columnwidth]{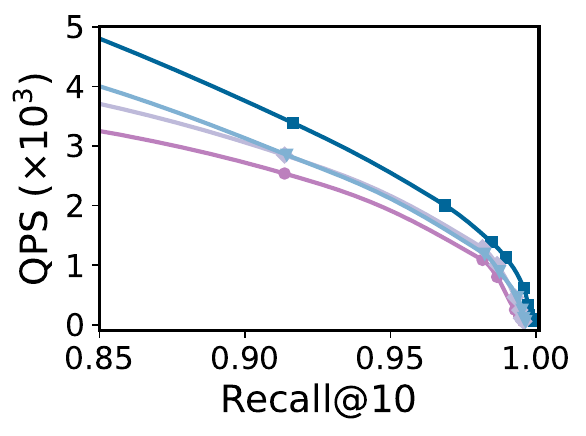}
}
\vspace{-3mm}
\caption{Ablation study for pair edge selection (Exp. 3)}

\vspace{-4mm}
\label{fig:ablation-region}
\end{figure}

\noindent\underline{Exp. 3-3: Offline Pair Edge Selection.}
Fig.~\ref{fig:ablation-region} compares \ours with \ding{192} no hierarchy, which removes hierarchy-derived region pairs; \ding{193} no quotient graph, which removes quotient-derived region pairs; and
\ding{194} no smoothing, which retains a geometry-only hierarchy.

At the $0.99$ recall target, \ours achieves $1.09$--$2.43$x the QPS of these variants across the three datasets.
These results support combining hierarchy-derived spatial candidates with candidates lacking short quotient paths, and using graph smoothing to incorporate local connectivity into the partition.
Index sizes remain nearly unchanged under the same degree budgets.
For \ding{192}, \ding{193}, and \ding{194}, respectively, construction takes
$3.176$, $4.480$, and $4.281$\,s on \yaleb;
$47.928$, $69.075$, and $63.310$\,s on \multipie; and
$509.614$, $598.292$, and $565.587$\,s on \SIFT.
Removing hierarchy candidates yields the largest construction-time reduction, while omitting quotient candidates or smoothing provides smaller savings relative to the full method.

\begin{figure}

\subfigure[$\Mpair$]
{
    \includegraphics[width=0.3\columnwidth]{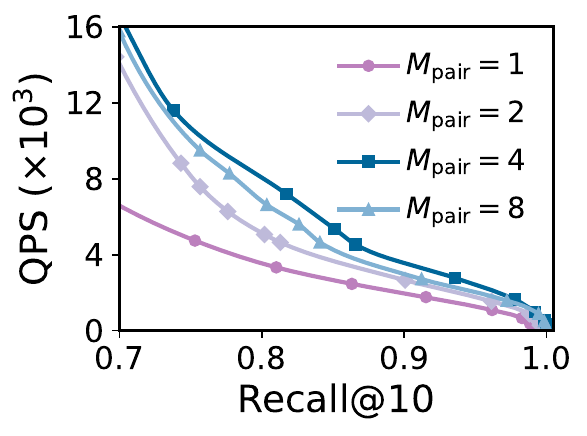}
    \label{fig:parameter-sensitive-mpair}
}
\hfill
\subfigure[{$b$}]
{
    \includegraphics[width=0.3\columnwidth]{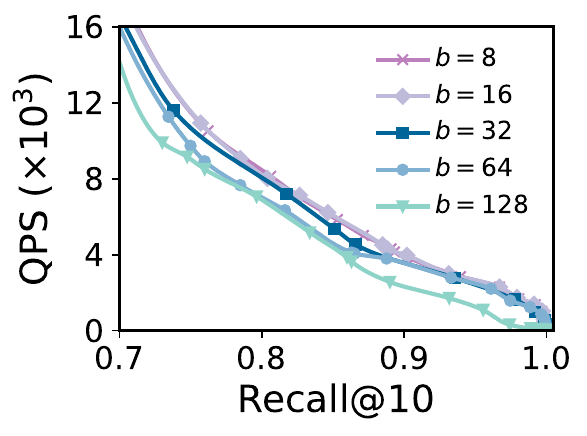}
    \label{fig:parameter-sensitive-b}
}
\hfill
\subfigure[{$t_s$}]
{
    \includegraphics[width=0.3\columnwidth]{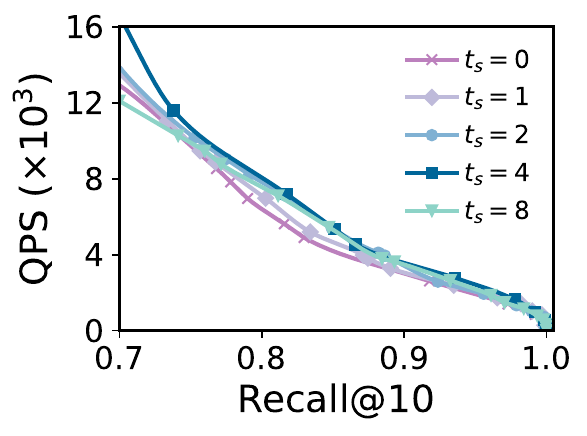}
    \label{fig:parameter-sensitive-t_s}
}

\subfigure[{$s_h$}]
{
    \includegraphics[width=0.3\columnwidth]{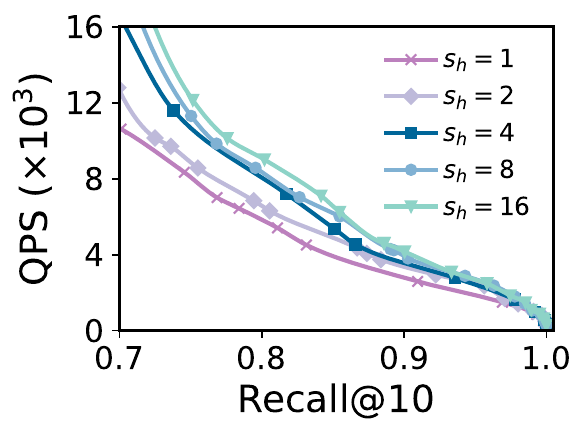}
    \label{fig:parameter-sensitive-s_h}
}
\hfill
\subfigure[{$\kappa$}]
{
    \includegraphics[width=0.3\columnwidth]{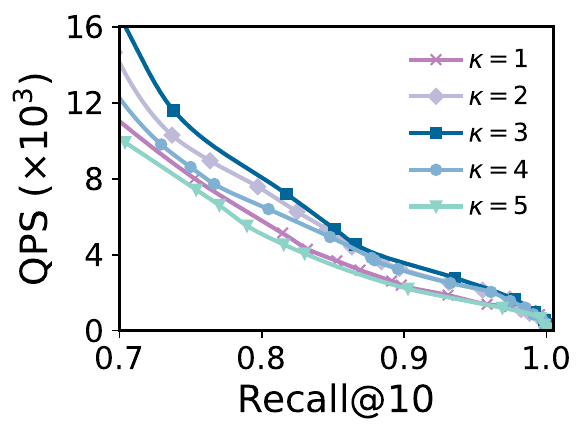}
    \label{fig:parameter-sensitive-kappa}
}
\hfill
\subfigure[{$r$}]
{
    \includegraphics[width=0.3\columnwidth]{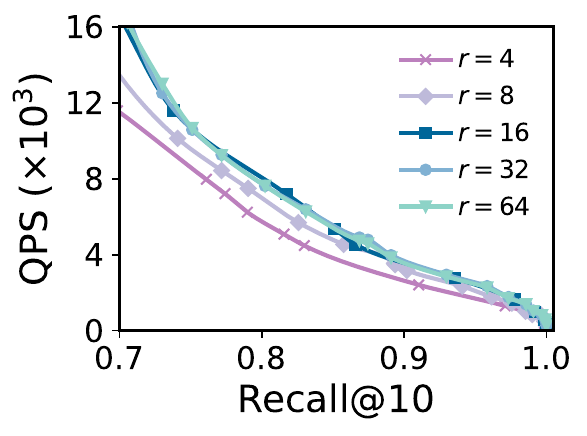}
    \label{fig:parameter-sensitive-r}
}

\vspace{-3mm}
\caption{Parameter sensitivity study on \yaleb (Exp. 4)}

\vspace{-4mm}
\label{fig:parameter-sensitive}
\end{figure}

\stitle{Exp. 4: Parameter Sensitivity.}
Fig.~\ref{fig:parameter-sensitive} examines $\Mpair$, $b$, $t_s$,
$s_h$, $\kappa$, and $r$ on \yaleb with $k=10$.
We vary one parameter at a time while keeping the other
parameters at their default values.
\textbf{(1) Pair-edge degree $\Mpair$:}
We vary $\Mpair$ over $\{1,2,4,8\}$ while keeping the total degree
limit $M$ fixed.
As shown in Fig.~\ref{fig:parameter-sensitive-mpair}, at a recall target of $0.99$, $\Mpair=4$ provides $4.23$x, $1.98$x, and $1.10$x QPS of $\Mpair=1,2,8$, respectively.
This supports allocating a moderate number of edges to inter-region
navigation. Too few pair edges limit the available entrances, whereas
a larger budget can displace useful original connections under the
fixed total degree limit.
\textbf{(2) Leaf capacity $b$:}
We vary $b$ over $\{8,16,32,64,128\}$, and their reported construction times are $5.09$, $4.97$, $4.96$,
$4.91$, and $4.82$\,s, respectively.
As shown in Fig.~\ref{fig:parameter-sensitive-b}, smaller leaves create more regions to process, but lead to larger index time.
The effect on query performance is more substantial.
At the $0.99$ recall target, $b=16$ achieves $1{,}394.57$ QPS,
compared with $977.77$ QPS for $b=32$ and $73.00$ QPS for
$b=128$.
Thus, leaf capacity affects the resolution of region selection and
its recall--QPS tradeoff.
\textbf{(3) Smoothing rounds $t_s$:}
We test $t_s\in\{0,1,2,4,8\}$.
As shown in Fig.~\ref{fig:parameter-sensitive-t_s}, at recall $1.0$, four rounds are $1.79$x as fast as no smoothing and $13.41$x as fast as eight rounds at this operating point.
These results support a finite smoothing budget: graph averaging can improve the partition, whereas excessive averaging can weaken the initial geometric distinctions.
\textbf{(4) Sampling budget $s_h$:}
We vary $s_h\in\{1,2,4,8,16\}$. Fig.~\ref{fig:parameter-sensitive-s_h} shows that increasing $s_h$ may lead to better search performance while it also increases the index building time.
The construction times increase from $4.02$ to $4.47$, $4.96$, $5.24$, and $5.50$\,s.
The default $s_h=4$ balances this construction cost with strong high-recall performance; larger budgets can help at other operating points but do not consistently dominate it.
\textbf{(5) Quotient-hop radius $\kappa$:}
We vary $\kappa$ over $\{1,2,3,4,5\}$. As shown in Fig.~\ref{fig:parameter-sensitive-kappa}, at the $0.99$ recall, the default $\kappa=3$ is $1.16$x,
$1.44$x, $1.14$x, and $1.50$x as fast as the other settings at this target.
A small hop radius admits destinations still connected through short multi-hop routes, whereas a larger radius excludes more destinations and can restrict the available quotient candidates.
The results support an intermediate exclusion radius.
\textbf{(6) Retained region partners $r$:}
We vary $r$ over $\{4,8,16,32,64\}$. As shown in Fig.~\ref{fig:parameter-sensitive-r}, at the $0.99$ recall target, increasing $r$ from $4$ to $8$ and $16$ raises QPS from $752.47$ to $846.47$ and $977.77$.
Further increasing it to $32$ and $64$ yields $1{,}052.76$ and $1{,}018.01$ QPS.
Increasing $r$ first leads to higher QPS, while the clearest saturation occurs between $r=32$ and $64$, QPS decreases by $3.30\%$ relative to $r=32$.
The total number of selected region pairs increases only $0.28\%$, although the partner cap doubles, indicating limited additional selections without further QPS improvement.

\begin{figure}

\subfigure[QPS]
{
    \includegraphics[width=0.3\columnwidth]{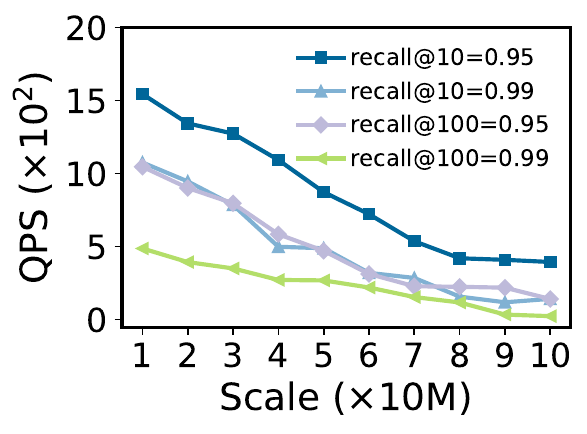}
    \label{fig:scal-qps}
}
\hfill
\subfigure[Index size]
{
    \includegraphics[width=0.3\columnwidth]{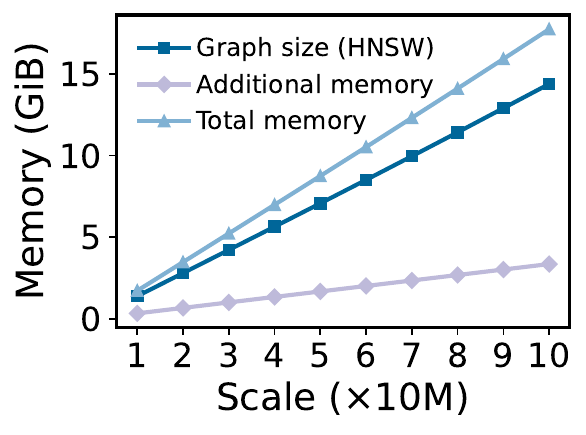}
    \label{fig:scal-size}
}
\hfill
\subfigure[Index time]
{
    \includegraphics[width=0.3\columnwidth]{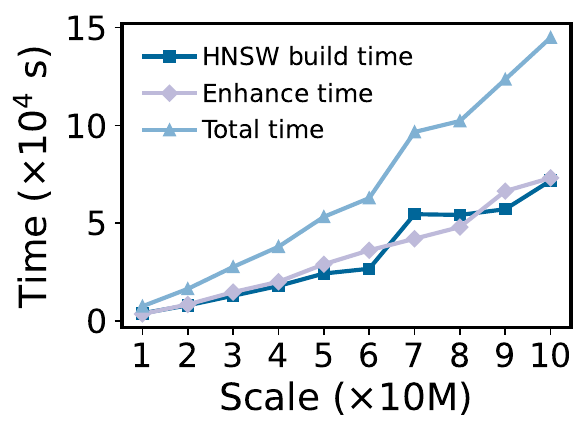}
    \label{fig:scal-time}
}

\vspace{-3mm}
\caption{Scalability evaluation (Exp. 5)}

\vspace{-4mm}
\label{fig:scalability}
\end{figure}

\stitle{Exp. 5: Scalability.}
We evaluate the scalability of \ours on \kw{SIFT100M}~\cite{sift} by scaling the dataset size from 10M to 100M points.
Fig.~\ref{fig:scal-qps} reports QPS achieved at fixed target recalls. As the scale increases, query throughput consistently decreases for all settings, while the method still delivers 394.3 QPS at Recall@10 = 0.95 and 140.8 QPS at Recall@100 = 0.95 on the 100M dataset.
The index size (excluding base vectors) curves in Fig.~\ref{fig:scal-size} exhibit nearly linear growth with data scale for both the underlying HNSW graph and our enhancements.
Fig.~\ref{fig:scal-time} shows that both HNSW build time and the enhancement time increase with scale, and together lead to a total build time of 7.49k s at 10M and 144.90k s at 100M. The enhancement stage is comparable to HNSW construction cost across the full range, which confirms that the overall index construction remains practical even at the largest scale tested.

\section{Conclusion}
\label{sec:con}

In this paper, we study the approximate line nearest neighbor search in vector databases. We propose \ours, which enhances a proximity graph with selected pair edges while retaining the original vector dimension and total degree limit.
Our extensive experiments demonstrate that \ours outperforms all existing methods for LANN search.
At Recall@10 of 0.99, \ours\ delivers up to 4.23x QPS of \DirSearch, and a higher recall that \DirSearch cannot reach. Compared with \Lift, it achieves better search performance while using only 1.86\% of the index space and 2.15\% of the construction time on full datasets on average. \ours achieves index construction that is more than 3 orders of magnitude faster, with a much smaller index size and higher search performance than \Theory.

\balance
\bibliographystyle{ACM-Reference-Format}
\bibliography{reference}

\end{document}